\documentclass[11pt,preprint,preprintnumbers,amsmath,amssymb]{revtex4}

\usepackage{graphicx}
\usepackage{dcolumn}
\usepackage{bm}

\usepackage{epsfig}

\def\iso#1#2{\mbox{${}^{#2}{\rm #1}$}}
\def\he#1{\iso{He}{#1}}
\def\li#1{\iso{Li}{#1}}
\def\be#1{\iso{Be}{#1}}
\def\b#1{\iso{B}{#1}}

\def\pfrac#1#2{\left( \frac{#1}{#2} \right)}

\newcommand\beq{\begin{equation}}
\newcommand\eeq{\end{equation}}
\newcommand\beqar{\begin{eqnarray}}
\newcommand\eeqar{\end{eqnarray}}
\def\la{\mathrel{\mathpalette\fun <}}
\def\ga{\mathrel{\mathpalette\fun >}}
\def\fun#1#2{\lower3.6pt\vbox{\baselineskip0pt\lineskip.9pt
  \ialign{$\mathsurround=0pt#1\hfil##\hfil$\crcr#2\crcr\sim\crcr}}}

\newcommand\EE[2]{{#1}\!\times\! 10^{#2}}

\newcommand\nnu{\mbox{$N_{\rm \nu,eff}$}}

\begin{document}
\title{Primordial Nucleosynthesis for the New Cosmology: \\ Determining Uncertainties and Examining Concordance}
\author{Richard H. Cyburt}
\affiliation{Department of Physics, University of Illinois, Urbana, IL 61801, USA \\ TRIUMF, Vancouver, B.C. V6T 2A3, Canada}
\email{cyburt@triumf.ca}
\begin{abstract}
Big bang nucleosynthesis (BBN) and the cosmic microwave background
(CMB) have a long history together in the standard cosmology. BBN
accurately predicts the primordial light element abundances of
deuterium, helium and lithium.  The general concordance between the
predicted and observed light element abundances provides a direct
probe of the universal baryon density.  Recent CMB anisotropy
measurements, particularly the observations performed by the WMAP
satellite, examine this concordance by independently measuring the
cosmic baryon density. Key to this test of concordance is a
quantitative understanding of the uncertainties in the BBN light
element abundance predictions.  These uncertainties are dominated by
systematic errors in nuclear cross sections, however for helium-4 they
are dominated by the uncertainties in the neutron lifetime and
Newton's G.  We critically analyze the cross section data, producing
representations that describe this data and its uncertainties, taking
into account the correlations among data, and explicitly treating the
systematic errors between data sets.  The procedure transforming these
representations into thermal rates and errors is discussed.  Using
these updated nuclear inputs, we compute the new BBN abundance
predictions, and quantitatively examine their concordance with
observations.  Depending on what deuterium observations are adopted,
one gets the following constraints on the baryon density: $\Omega_{\rm
B}h^2=0.0229\pm0.0013$ or $\Omega_{\rm B}h^2 =
0.0216^{+0.0020}_{-0.0021}$ at 68\% confidence, fixing
$N_{\nu,eff}=3.0$.  If we instead adopt the WMAP baryon density, we
find the following deuterium-based constraints on the effective number
of neutrinos during BBN: $N_{\nu,eff}=2.78^{+0.87}_{-0.76}$ or
$N_{\nu,eff}=3.65^{+1.46}_{-1.30}$ at 68\% confidence. Concerns over
systematics in helium and lithium observations limit the confidence
constraints based on this data provide.  BBN theory uncertainties are
dominated by the following nuclear reactions: $d(d,n)\he3$, $d(d,p)t$,
$d(p,\gamma)\he3$, $\he3(\alpha,\gamma)\be7$ and $\he3(d,p)\he4$.
With new nuclear cross section data, light element abundance
observations and the ever increasing resolution of the CMB anisotropy,
tighter constraints can be placed on nuclear and particle
astrophysics.
\end{abstract}
\maketitle

\section{Introduction}

The field of cosmology has recently entered a golden age.  An age
where a global picture of the universe is crystallizing because of new
precision observations that can test the basic framework of the
standard cosmological model.  With the plethora of new data, it is
important to review and test the fundamental theoretical pillars of
cosmology.  These pillars are the theory of general relativity and the
universal expansion, big bang nucleosynthesis (BBN), and the relic
cosmic background radiation.

\subsection{History}

Knowledge of general relativity and the discovery in 1929 by
Hubble that the universe was possibly expanding \cite{hubble}, led to
the idea that one could extrapolate backwards and conclude that the
universe was hotter and denser in the past.  This idea became what is
currently called the ``hot big bang'' model of the universe.  Almost
20 years later it was realized that at early enough times, the
universe would have been hot and dense enough for nuclear fusion to
take place.  This epoch of primordial nucleosynthesis could explain
the large abundances of hydrogen and helium seen in the universe (and
ultimately the trace D, \he3, and \li7 abundances), first explored by
Gamow (1946), Alpher, Bethe and Gamow (1948), Hayashi (1950), and
Alpher, Follin and Herman (1953)
\cite{gamow,abg,hyashi,afh}. 

The ``hot big bang'' model also predicted a relic photon background,
created when ions recombined with electrons to form neutral atoms
(Alpher \& Herman, 1948,1949 \cite{ah1,ah2}).  In 1965, this uniform 3
Kelvin background was detected by Penzias and Wilson for the first
time in the microwave band
\cite{penziaswilson}. This cosmic microwave background (CMB) offered
supporting evidence for the ``hot big bang'' model and stimulated further
refinements in the theory of big bang nucleosynthesis (Peebles, 1966;
Wagoner, Fowler \& Hoyle, 1967 \cite{peebles66,wfh67}).  

A decade ago, the COBE satellite detected for the first time the
$1:10^5$ intrinsic temperature fluctuations in the CMB
\cite{cobe}.  During the last five years,
many more CMB temperature anisotropy measurements have been made
(e.g. MAXIMA, BOOMERANG, DASI, CBI, ACBAR
\cite{max,boom,dasi,cbi,acbar}).  The latest of these observations
being from the WMAP satellite, with its first data release in early 2003
\cite{wmap}. 

These two pillars of cosmology offer a unique probe of early universe
physics; while their ultimate concordance depends upon the accuracy of
the standard cosmological model and of the observations driving this
precision era. These observations are so precise that we can test and
constrain cosmology in a profound and fundamental way.  For reviews of
BBN see Schramm and Wagoner (1979)~\cite{schwag}, Yang {\em et
al}. (1984)~\cite{yang}, Boesgaard and Steigman
(1985)~\cite{boessteig}, Kolb and Turner (1990)~\cite{kolbturner},
Walker {\em et al}. (1991)~\cite{wssok}, Sarkar (1996)
\cite{sarkar}, Olive, Steigman and Walker (2000)~\cite{osw}, Tytler 
{\em et al} (2000)~\cite{tytrev} and the Particle Data Group BBN
Review by Fields and Sarkar (2002)~\cite{pdg}.  For reviews of CMB
theory see White, Scott and Silk (1994)~\cite{whitescott}, Tegmark
(1995)~\cite{tegmark}, Van der Veen (1998)~\cite{vanderveen},
Kamionkowski and Kosowsky (1999)~\cite{kamkos} and Hu and Dodelson
(2002)~\cite{hudodl} and of CMB observations see Wang {\em et al}
(2002)~\cite{wang} for a pre-WMAP evaluation and the individual group
papers mentioned above.

\subsection{Goals}

Over the past decade, a major thrust of research in BBN has been
towards increasing the rigor of the analysis.  On the theory side, the
key innovation was to calculate the errors in the light element
predictions in a systematic and statistically careful way. This was
done using Monte Carlo analyses (Krauss and Romanelli
1990~\cite{kr90}; Smith, Kawano and Malaney 1993~\cite{skm}; Krauss
and Kernan 1995~\cite{kk95}; Hata {\em et al}. 1996~\cite{hata};
Fiorentini {\em et al}. 1998~\cite{flsv}; Nollett and Burles
2000~\cite{nb00}; Cyburt, Fields and Olive 2001~\cite{cfo1}; Coc {\em
et al}. 2002~\cite{cvcr}), which account for nuclear reaction
uncertainties and their propagation into uncertainties in the light
element abundance predictions.  These calculations are essential
because they allow for a careful statistical comparison of BBN theory
with observational constraints; in addition, they point the way toward
improvements in the theory calculation.

In its standard $N_\nu = 3.0$ form, primordial nucleosynthesis is a
one parameter theory, depending only on the baryon-to-photon ratio
$\eta\equiv n_{\rm B}/n_\gamma$.  This is related to the cosmic baryon
density; assuming that H and \he4 are the dominant constituents after
the epoch of BBN ($\rho_{\rm B}=n_{\rm H}m_{\rm H}+n_{\rm He}m_{\rm
He}$), yields the relation:
\beq
\label{eqn:baryeta}
273.66\Omega_{\rm B}h^2 \equiv 10^{10}\eta \ [1.0 -
0.0071186Y_p]\left( \frac{{\rm G}_N}{6.673\times 10^{-8} {\rm \
cgs}}\right) \left( \frac{T_{\gamma,0}}{2.725 {\rm \ K}}\right)^3
\eeq
where $\Omega_{\rm B}$ is the current baryon density relative to the
critical density, $\rho_c\equiv 3H^2/8\pi{\rm G}_N$.  $H$ is the
current Hubble parameter, usually defined as $H=100h {\rm \ km \
s^{-1}
\ Mpc^{-1}}$ and G${}_N$ is Newton's gravitational constant.  $Y_p$ is
the primordial, post-BBN mass fraction of baryons in the form of \he4
and $T_{\gamma,0}$ is the current temperature of the cosmic microwave
background.  Since the mass of the proton is not the same as the mass
per baryon of \he4, $Y_P$ appears in eqn.~\ref{eqn:baryeta}.  One can
see that with the convolution of BBN theory predictions with light
element observations, constraints on the baryon density can be placed.
The agreement between the various baryon density constraints from
different light element observations places quantitative limits on
their concordance.  Deviations from concordance, suggests unknown
observational, experimental or theoretical systematics.  The latter
possibly indicating the need for new physics in the standard BBN
framework.  This has been extensively explored, the reader is
recommended the following incomplete list of
reviews~\cite{sarkar,ssg,dolgov}.  With little change in observational
or experimental data, these bounds have remained relatively unchanged
over the last few years.

The recent boon in CMB anisotropy measurements, offers to reshape the
cosmological landscape.  What these observations bring to the table is
an independent measure of the cosmic baryon density.  This independent
measurement of the baryon content examines the general concordance of
the BBN light element abundance theory predictions and their observed
values, and tests the basic framework of the hot big bang model.  It
acts as a ``tie-breaker'' for the various light element
observation-based baryon density
constraints~\cite{st98,jkks,flsv,bnt,nb00,cfo1,cvcr}.

Key to this test, is an understanding of the dominant uncertainties in
the light element predictions.  These uncertainties stem from the
systematic errors in nuclear cross sections.  We present a new
procedure for determining cross section representations and their
uncertainties and describe how they propagate into thermal rates and
the light element predictions.  With this updated nuclear network, we
then quantify the concordance between the light element abundance
observations and their predictions, and the CMB.  With this level of
concordance set for the standard cosmological model, we can test and
constrain non-standard models.  We use primordial nucleosynthesis and
the cosmic microwave background together to probe early universe
physics spanning times from 1 sec to 400,000 yrs after the big bang
and beyond.  This work follows naturally from the work performed by
Cyburt, Fields and Olive (2001,2002,2003)~\cite{cfo1,cfo2,cfo3} and
continues with the same guard as the research by Smith, Kawano and
Malaney (1993)~\cite{skm}, and Nollett and Burles (2000)~\cite{nb00}.

This paper is organized as follows: In $\S$~\ref{sect:formalism}, we
describe the formalism of creating representations and uncertainties
for cross sections and transforming them into thermal rates and
uncertainties.  In $\S$~\ref{sect:results}, we discuss the resulting
cross sections and thermal rates and their impact on the light element
predictions of primordial nucleosynthesis.  We then establish the
level of concordance existing between light element observations,
their predictions and the CMB in $\S$~\ref{sect:disc}, followed by conclusions in $\S$~\ref{sect:concl}.


\section{Formalism}
\label{sect:formalism}
In this new age of precision cosmology, it is increasingly important
to have an up-to-date and accurate theory of primordial
nucleosynthesis.  Since BBN's uncertainties stem from uncertainties in
nuclear cross section data, we develop here a rigorous and
reproducible procedure for determining accurate representations of
that data.  There are several requirements we wish to impose on this
analysis. (1.)  The representation of the data must be model
independent, other than basic assumptions of functional form and
demanding sufficient smoothness. (2.)  The treatment should be global,
all data is analyzed simultaneously, avoiding operator's discretion
and so-called ``chi-by-eye'' systematics.  (3.)  There should be
explicit treatment of a.) the correlations among data in a data set
and b.) the discrepancies between different data sets' normalizations.
These explicit and implicit normalization errors dominate over the
statistical uncertainties in the data.

With these goals in mind, we set out to build a framework for
representing cross section data for the nuclear reactions important
for an accurate BBN calculation, seen in table \ref{tab:rates}.  To
begin, we will discuss the way cross section data is presented,
defining notation that will be useful.  We then present the scheme for
determining the best representation of data and the uncertainty in
such a representation.  Finally, presenting the reactions most
important for primordial nucleosynthesis, and their fits and
uncertainties.

\begin{table}[ht]
\caption{Shown in this table are the 12 most important reactions 
affecting the predictions of the light element abundances (\he4, D,
\he3, \li7). }
\label{tab:rates}
\begin{tabular}{ll}
\hline\hline
Reactions\\
\hline
$n$-decay                \\
$p(n,\gamma)d$           \\
$d(p,\gamma)\he3$        \\
$d(d,n)\he3$             \\
$d(d,p)t$              \\
$\he3(n,p)t$             \\
$t(d,n)\he4$             \\
$\he3(d,p)\he4$          \\
$\he3(\alpha,\gamma)\be7$\\
$t(\alpha,\gamma)\li7$   \\
$\be7(n,p)\li7$          \\
$\li7(p,\alpha)\he4$     \\
\hline\hline
\end{tabular}
\end{table}

\subsection{Data Sets}
Ideally, a cross section datum contains four numbers: expectation
values and uncertainties for the cross section and energy.
Uncertainties in the energy are typically, negligibly small.  One
difficulty in measuring cross sections is determining their absolute
normalization.  In addition to the statistical error in each point a
normalization uncertainty is assigned for a particular data set.  In
many cases this systematic normalization uncertainty dominates over
the the statistical.  When using cross section data, we must take into
account the fact that data from a particular data set are correlated
with each other due to this normalization error.  To help with
visualization and to find the correlation matrix, we define a random
variable to draw from to produce a data point:
\beq
\underline{x}_i = (1+\epsilon\underline{z}_0)(\mu_i +
\sigma_i\underline{z}_i).
\eeq
We denote a random variable by underlining it (e.g. $\underline{x}_i$,
$\underline{z}_0$, $\underline{z}_i$).  The random variables,
$\underline{z}_0$ and $\underline{z}_i$, are assumed to be
uncorrelated random variables, with zero mean and unit variance.
Notice that the mean normalization is unity, we could have allowed
another normalization, but opted not to because we do not have a
reference point to normalize to.  In principle, one could use theory
to determine an experiment's normalization, but we choose the
model-independent approach, relying on the data as is.  Not
renormalizing here will lead us to the separate treatment of
systematic differences between data sets, and the assignment of an
overall ``theory'' normalization uncertainty.  The expectation values
and correlation matrix elements are:
\beqar
{\rm Exp}[\underline{x}_i] &=& \langle \underline{x}_i \rangle = \mu_i \\
{\rm Cov}[\underline{x}_i, \underline{x}_j] &=& \langle
\underline{x}_i\underline{x}_j\rangle - \langle
\underline{x}_i\rangle\langle\underline{x}_j\rangle =
(1+\epsilon^2)
\sigma_i^2\delta_{ij} + \epsilon^2\mu_i\mu_j
\eeqar
Generalizing this for multiple data sets we get,
\beq
{\mathcal C}_{i_n,j_n} = (1+\epsilon_n^2)
\sigma_{i_n}^2\delta_{{i_n}{j_n}} +
\epsilon_n^2\mu_{i_n}\mu_{j_n}
\eeq
where $i_n$ denotes the $i^{th}$ data point in the $n^{th}$ data set.
The inverse covariance matrix is:
\beq
{\mathcal C}^{-1}_{i_n,j_n} =
\frac{\delta_{{i_n}{j_n}}}{(1+\epsilon_n^2)\sigma_{i_n}^2} -
\frac{\frac{\epsilon_n^2
\mu_{i_n}\mu_{j_n}}{(1+\epsilon_n^2)^2\sigma_{i_n}^2\sigma_{j_n}^2} }{1
+ \frac{\epsilon_n^2}{(1+\epsilon_n^2)}\sum_{k_n}
\left(\frac{\mu_{k_n}}{\sigma_{k_n}}\right)^2}.
\eeq
It is this inverse covariance matrix that will be used in the later
best fit calculation.  In the case where the normalization error is
small, the covariance matrix reduces to the standard diagonal form
with the statistical errors as the diagonal elements.  In the case
where the normalization errors dominate, the inverse matrix becomes:
\beq
{\mathcal C}^{-1}_{i_n,j_n} =
\frac{1}{(1+\epsilon_n^2)\sigma_s^2}\left( \delta_{{i_n}{j_n}} -
\frac{
\mu_{i_n}\mu_{j_n} }{\sum_{k_n}
\mu_{k_n}^2}\right),
\eeq
where $\sigma_s$ is a typical, albeit small statistical uncertainty.
When data sets are large, the second term in the parentheses becomes
small, thus the covariance matrix again reduces to the standard form
with slightly inflated statistical errors.  Thus, one can see that
quite generally, the statistical uncertainty is the dominant
contribution to the inverse covariance matrix, not the total
uncertainty.  When data sets are small, the covariance matrix is
highly non-diagonal.  However, since our prescription combines several
data sets, one is often significantly larger than the others, thus
smaller data sets will have less impact on the fit.

If we were dealing with one data set, we would not necessarily need
this formalism.  As noted in D'Agostini (1994) \cite{dagostini}, it is
generally better to treat the normalization error separate from the
statistical one (e.g.  determining a best fit based on the statistical
uncertainties alone and adding in the normalization error after the
fitting process), however we would be ignoring correlations.  Since we
are combining multiple data sets in a meta-analysis, we must include
the normalization correlations between data points as well as find an
effective overall normalization error to add after the fitting
process. Different data sets may disagree on the shape, or cover
different energy regions.  Therefore, we will continue with the
formalism we have laid out.

\subsection{Creating Representations}
We are interested in determining the best fit parameters for some
general, linear parameterization of the data.  In each data set, $n$
we have the following data; a position variable (e.g. the energy)
$x_{i_n}$, the expectation value of the function (e.g. the cross
section or $S$-factor), $y_{i_n}$ measured at $x_{i_n}$, and the covariance between
data points ${\mathcal C}_{i_n,j_n}$.  We will assume that each data
set is independent from all others.

In the standard treatment, we determine the best fit parameters by
minimizing a $\chi^2$.  For simplicity, we choose a linear combination
of known functions for our parameterization, $y(x) = \sum_p a_p {\tt
X}_p(x)$, where $a_p$ and ${\tt X}_p(x)$ are the $p^{\rm th}$ of $P$
fitting parameters and fitting functions evaluated at $x$. For
example, a polynomial fit (which we will adopt) has ${\tt X}_p(x) =
x^p$.  To begin, I will look at the case for one data set.  We define:
\beq
\chi^2 = \sum_{i,j=1}^I {\mathcal C}^{-1}_{i,j} \left[ \sum_{p=1}^P
a_p{\tt X}_p(x_i) - y_i 
\right] \left[ \sum_{q=1}^P a_q{\tt X}_q(x_j) - y_j \right];
\eeq
We reiterate here that a calligraphic ${\mathcal C}$ denotes the
covariance between data points.

When determining the best fit parameters by minimizing $\chi^2$, we
can re-write it as:
\beq
\chi^2 = \chi^2_{\rm min} +
\sum_{p,q=1}^P C^{^{-1}}_{p,q} (a_p - \hat{a}_p) (a_q - \hat{a}_q),
\eeq
where $\hat{a}_p$ and $C^{^{-1}}_{p,q}$ are the most likely values and
the inverse covariance between the $p^{\rm th}$ and $q^{\rm th}$
parameters.  We note here that an italic $C$ denotes the covariance
between fitting parameters, not the data points.  The best fit and its
variance are then
\beqar
\mu(x) &=& \sum_p \hat{a}_p {\tt X}_p(x) \\
\sigma^2(x) &=& \sum_{p,q} C_{p,q} {\tt X}_p(x) {\tt X}_q(x).
\eeqar

The most likely parameter values are  given by:
\beq
\label{eqn:mlv}
\hat{a}_p = \sum_{q=1}^P C_{p,q} {\mathcal A}_q
\eeq
where
\beqar
{\mathcal A}_q &=& \frac{1}{2} \sum_{i,j} {\mathcal C}^{-1}_{i,j}
\left[ {\tt X}_q(x_i)y_j + y_i{\tt X}_q(x_j) \right], \\
C^{^{-1}}_{p,q} &=& \frac{1}{2} \sum_{i,j} {\mathcal C}^{-1}_{i,j}
\left[ {\tt X}_p(x_i){\tt X}_q(x_j) + {\tt X}_q(x_i){\tt X}_p(x_j) \right].
\eeqar
Since were are demanding linear fitting functions, the $\chi^2$ is
quadratic in the fitting parameters, thus yielding a correlated
gaussian probability distribution with the form:
\beq
{\mathcal L}(\vec{a}) = \frac{\exp{\left[ -\frac{1}{2}
(\vec{a}-\vec{\hat{a}})^{^T}\!\!\cdot {\bf C}^{^{-1}}\!\!\cdot (\vec{a} -
\vec{\hat{a}}) \right]}}{\sqrt{(2\pi)^P \,{\rm det}({\bf C})}}. 
\eeq

When generalizing this to more than one data set, we have to ask
ourselves how do we want to weight the data and each data set.  If we
wanted to rely strictly on the data itself, then the $\chi^2$ is
simply the sum of the $\chi^2$'s from each experiment.  This in turn
propagates into the fitting parameter likelihood distribution as:
\beq
{\mathcal L}(\vec{a}) = \prod_{n=1}^N \frac{\exp{\left[
-\frac{1}{2} (\vec{a}-\vec{\hat{a}}_n)^{^T}\!\!\cdot {\bf
C}^{(n)^{-1}}\!\!\cdot (\vec{a} - \vec{\hat{a}}_n)
\right]}}{\sqrt{(2\pi)^P \,{\rm det}({\bf C}^{(n)})}}.  
\eeq

This scheme gives more weight to the data sets with more data points,
with ${\rm det}({\bf C}^{(n)})$ scaling like $1/{\rm I}_n$, where
${\rm I}_n$ is the number of data points in the $n^{\rm th}$ of $N$
data sets.

If instead we wanted to treat data sets on an equal footing, then the
parameter likelihood distribution takes on the form:
\beq
\label{eqn:Lnouse}
{\mathcal L}(\vec{a}) = \frac{1}{\rm N}\sum_{n=1}^N
\frac{\exp{\left[ -\frac{1}{2} (\vec{a}-\vec{\hat{a}}_n)^{^T}\!\!\cdot
{\bf C}^{(n)^{-1}}\!\!\cdot (\vec{a} - \vec{\hat{a}}_n)
\right]}}{\sqrt{(2\pi)^P \,{\rm det}({\bf C}^{(n)})}}, 
\eeq
Notice the products of likelihoods has been replaced with a sum of
likelihoods in this non-standard treatment.  We can see with this
likelihood, that a $\chi^2$ analysis becomes more complicated.  The
effective $\chi^2 = -2\ln{\mathcal L}$ is no longer quadratic in the
fitting parameters, thus making the distribution non-gaussian and
possibly multi-peaked.
 
Since we are not only determining the magnitude, but the shape of a
function, we should rely more on the data sets that have more points.
Thus the first prescription is appropriate for our purposes.  The
minimum $\chi^2$ in this prescription is:
\beq
\chi^2_{\rm min} = \sum_{n=1}^N \chi^2_{{\rm min},n} +
\sum_{n=1}^N (\vec{\hat{a}}-\vec{\hat{a}}_n)^{^T}\!\!\cdot {\bf
C}^{(n)^{-1}}\!\!\cdot (\vec{\hat{a}} - \vec{\hat{a}}_n).
\eeq
The best
fit parameters are  still given by equation \ref{eqn:mlv}, but where
\beqar
{\mathcal A}_q &=& \frac{1}{2} \sum_{n=1}^N\sum_{i_n,j_n=1}^{I_n}
{\mathcal C}^{-1}_{i_n,j_n} 
\left[ {\tt X}_q(x_{i_n})y_{j_n} + y_{i_n}{\tt X}_q(x_{j_n}) \right], \\
C^{^{-1}}_{p,q} &=& \frac{1}{2} \sum_{n=1}^N \sum_{i_n,j_n=1}^{I_n}
{\mathcal C}^{-1}_{i_n,j_n} 
\left[ {\tt X}_p(x_{i_n}){\tt X}_q(x_{j_n}) + {\tt X}_q(x_{i_n}){\tt
X}_p(x_{j_n}) \right]. 
\eeqar
Note that if data sets disagree, the minimum $\chi^2$ per degree of
freedom ($\chi^2_\nu=\chi^2/\nu$) will be large, where $\nu$ is the
number of degrees of freedom.  With the covariance in the fitting
parameters depending solely on the covariance among the data, which as
discussed earlier depends mainly on the statistical uncertainty, the
error in the mean, $\sigma(x)$, is a measure of the statistical
uncertainty only.  When we have a lot of data, this error will be
small due to the $1/\sqrt{N}$ suppression of the error in the mean.
Thus if we have two data sets with a large quantity of data, but both
systematically offset from each other, the error will be
underestimated.

This procedure does not take into account the systematic differences
between data sets.  There are various ways of treating uncertainty
assignment with discrepant data.  The Particle Data Group
prescription, is to blow up the error in the mean by the factor
$\sqrt{\chi^2_\nu}$ \cite{pdg}.  This has the virtue that it does take
into account systematic differences and effectively forces the
$\chi^2_\nu$ to be unity.  Its limitation lies in the fact that this
scale factor does not cancel out the $1/\sqrt{N}$ suppression in the
error, thus for sufficiently large data sets, this error assignment
will still underestimate the true errors when using two discrepant
data sets.

When dealing with a one parameter fit or renormalization where
systematics dominate, Cyburt, Fields and Olive (2001) showed that the
appropriate scale factor for discrepant data is $\sqrt{\chi^2}$, not
$\sqrt{\chi^2_\nu}$
\cite{cfo1}.  This error assignment turns out to be the weighted
dispersion about the mean:
\beq
\sigma^2 = \frac{\sum_i \left( \frac{y_i - \mu}{\sigma_i}
\right)^2}{\sum_i \frac{1}{\sigma_i^2}}.
\eeq
This approach reproduces well the uncertainties when discrepant data
are present and has the added virtue that it continues to minimize the
variance and does not scale with the number of data points.
Ultimately this error is a measure of agreement between data sets.  If
agreement is met, then the error in the mean will dominate over this
error.  The limitation of both these methods, is that they do not
treat the error as if it varies with respect to the position variable
$x$ (e.g. energy).  However, if the differences between data sets is
attributable to an unknown normalization error, then assuming no
energy dependence is appropriate.

The Nollett and Burles (2000) \cite{nb00} compilation does not
explicitly calculate systematic uncertainties.  They create samples of
mock data, including the intrinsic normalization errors, and adopt
piecewise, smooth B-spline representations of cross sections, dividing
the energy range into smaller bins.  Each realization is thermally
averaged and propagated through the BBN calculation.  This treatment
has the virtue that it has an explicit treatment of the normalization
errors, however their B-spline fitting procedure does not take into
account the correlations between data points.  Also, this method
simply blows up the errors by reducing the number of points
contributing to the fit in a particular energy bin.  This method's
main limitation is that it introduces some arbitrariness into where
the data cuts are placed, and that it is still dealing with a strictly
statistical uncertainty and not a systematic one.  If discrepant data
exist such that it lies outside the typical error size, then the
Nollett \& Burles method will tend to underestimate the true
uncertainty.  In addition, the energy correlations of the cross
section data are not included (by assumption) in their fitting
procedure, thus affecting their best fit values.

It is clear that a procedure is needed to take into account the
systematic errors.  We will assume here, that the systematic errors
are purely normalization errors and as such are constant functions of
energy.  We will adopt the Cyburt, Fields and Olive 2001 \cite{cfo1}
sample variance as a measure of discrepant data.  Generalizing its
form to take into account the correlations between data, we get:
\beq
\delta_{disc}^2 \equiv \frac{\sum_n \sum_{i_n,j_n} {\mathcal
C}_{i_n,j_n}^{-1} \left[ \mu(x_{i_n}) - y_{i_n}\right] \left[
\mu(x_{j_n}) - y_{j_n}\right]}{\sum_n \sum_{i_n,j_n} {\mathcal
C}_{i_n,j_n}^{-1} \mu(x_{i_n}) \mu(x_{j_n})}.
\eeq
In addition, we need to calculate the normalization error inherent to
the data.  We choose a weighting scheme, such that data sets that
agree with the fit are given more weight than data sets that disagree,
since we have already taken into account discrepant data sets.  We
define the intrinsic normalization error to be:
\beq
\delta_{norm}^2 \equiv \frac{\sum_n
\frac{\epsilon_n^2}{\chi^2_n}}{\sum_n \frac{1}{\chi_n^2}}
\eeq
Here, $\chi_n^2$ is the minimum $\chi^2$ per datum of data set $n$,
given the best fit parameters.  The total normalization error is then
the quadrature sum of these two systematic errors $\delta^2 =
\delta_{disc}^2 + \delta_{norm}^2$.  This propagates into our final
error as:
\beq
\sigma^2(x) = (1+\delta^2)\sigma^2_{stat}(x) + \delta^2\mu(x)^2 = \sum_{p,q} \left[ (1+\delta^2) C_{p,q} + \delta^2
\hat{a}_p\hat{a}_q \right] {\tt X}_p(x){\tt X}_q(x).
\eeq

There is no unique way to assign a systematic uncertainty.  However,
any determination being, based on the same data, must agree with the
overall results of this prescription.  This leaves us with the
question of how can we further improve these uncertainties.  There are
two ways we can improve our errors, (1) we can get new, more accurate
and precise data and (2) we can, with sufficient reason, exclude data
sets, in an effort to remove the cause of the systematic errors.  As
there is not an un-biased way of performing the latter, we rely on the
former for the future progress of this type of analysis.  

\subsection{Thermal Averaging}

Thermonuclear reaction rates and the reaction networks they belong to,
play a key role in nuclear astrophysics theory, ranging from stellar
interiors, supernovae explosions to big bang nucleosynthesis.  A large
base of work has been done in this field.  Reaction rate formalism is
thoroughly reviewed in Clayton's ``Principles of Stellar Evolution and
Nucleosynthesis'' (1983) \cite{clayton} and Rolf and Rodney's
``Cauldrons in the Cosmos: Nuclear Astrophysics'' (1988)
\cite{cauldrons}.  Compilations of nuclear data and thermonuclear
rates began with the pioneering work of William Fowler
\cite{FCZI,FCZII,CF88}.  A recent update has been provided by the
NACRE collaboration \cite{nacre}.  Recent BBN rate compilations have
been performed by Smith, Kawano and Malaney \cite{skm}, Nollett and
Burles \cite{nb00} and the Cyburt, Fields and Olive \cite{cfo1}
tailored NACRE \cite{nacre} compilation.  

\subsection{Mapping Cross Sections into Thermal Rates}
\label{sect:3.1}

We want the rate at finite temperature, for 2-body interactions of the
type: $i+j\rightarrow k+l$, which is $\lambda_{i+j\rightarrow k+l}(T)
= N_A \langle \sigma_{i+j\rightarrow k+l}v\rangle$, where $N_A$ is
Avogadro's number.  The angle brackets denote thermal averages.  We
are interested in transforming energy dependent random functions into
temperature dependent random functions.  We define the transformation
from one to the other as:
\beq
\label{eqn:3-lambda}
\underline{\lambda}(T) = \int_0^\infty W(E,T) \underline{S}(E) dE,
\eeq
where $W(E,T)$ is a weighting function or kernal and
$\underline{S}(E)$ is the function we are transforming, either the
astrophysical $S$- or $R$- factor of a cross section.
$\underline{S}(E)$ depends on random variables (i.e. fitting
parameters), and thus is a random function, where the expectation
value is ${\rm Exp}[\underline{S}(E)] = \mu(E)$ and the variance is
${\rm Var}[\underline{S}(E)] =\sigma^2(E)$.

We want to know how this randomness propagates into
$\underline{\lambda}$.  The expectation value is:
\beq
\mu_\lambda(T) \equiv {\rm Exp}[\underline{\lambda}(T)] = \int_0^\infty W(E,T) \mu(E)
dE.
\eeq
The variance in $\lambda$ is then:
\beqar
\sigma_\lambda^2(T) \equiv {\rm Var}[\underline{\lambda}(T)]
\!\!&=&\!\! {\rm Exp}[\underline{\lambda}(T)^2] - {\rm
Exp}[\underline{\lambda}(T)]^2 \\
\!\!&=&\!\! \int_0^\infty\!\!\int_0^\infty W(E,T) W(E^\prime,T)
{\rm Cov}[\underline{S}(E),\underline{S}(E^\prime)] dE dE^\prime \\
&=&\!\! \int_0^\infty\!\!\int_0^\infty W(E,T) W(E^\prime,T)
\rho(E,E^\prime) \sigma(E)\sigma(E^\prime) dE
dE^\prime,
\eeqar
where $-1\leq\rho(E,E^\prime)\leq1$ is the energy dependent
correlation coefficient.  Notice that the variance depends on the
correlation of our random function between two energies.  If we
naively propagated the uncertainty as the transform of the standard
deviation:
\beq
\tilde{\sigma}_\lambda(T) = \int_0^\infty W(E,T) \sigma(E) dE,
\eeq
we would generally over-estimate the uncertainty, as seen in the
quadrature difference between these two error assignments.
\beq
\sigma_\lambda^2(T) - \tilde{\sigma}_\lambda^2(T) =
\int_0^\infty\!\!\int_0^\infty W(E,T) W(E^\prime,T)
[\rho(E,E^\prime)-1] \sigma(E)\sigma(E^\prime) dE  dE^\prime 
\eeq
Since $W(E,T)$ and $\sigma(E)$ are positive definite and the quantity
$\rho(E,E^\prime)-1 \leq 0$, the difference is always less than or
equal to zero.  Thus, inclusion of these energy correlations reduces
the total uncertainty in the thermal rates.  What form these rates and
errors take, depends on the what type of reaction we are dealing with
and how we have assigned systematic uncertainties.  Since we have
treated the systematic errors as normalization errors independent of
energy, it does not matter if we treat them in the integral or not.
Actually, performing the integral both with and without including the
systematic errors offers a nice way to double check the numerical
integration.  We now discuss the reactions important for primordial
nucleosynthesis and their fits based on the former procedure.


\section{Results}
\label{sect:results}

\subsection{Cross Sections}
\label{sect:cs_results}
Keeping in mind our efforts to maintain a rigorous and
model-independent analysis, we now implement this prescription for the
set of nuclear reactions that are most important for big bang
nucleosynthesis.  Along with the neutron lifetime and Newton's
G${}_N$, eleven key nuclear reactions dominate the uncertainties in
the BBN calculation of the light element abundances (Smith, Kawano and
Malaney 1993~\cite{skm}), determined by calculating the logarithmic
derivative of the predicted abundances with respect to each of the
reaction rates~\cite{flsv}.  Thus, the choice of nuclear compilation
with either its cross sections or thermal rates and their
uncertainties, will determine the accuracy of the final predictions.
The important work of Smith, Kawano and Malaney (1993) set a benchmark
and their error budget has been the standard for Monte Carlo work.
Nollett and Burles (2000) create their own compilation, but do not
present portable fits of their cross sections and thermal rates.  The
NACRE collaboration (Angulo {\em et al.}; 1999~\cite{nacre}),
represents a large effort to critically evaluate the available nuclear
data, presenting their adopted fits along with estimates of their
uncertainties.  Cyburt, Fields and Olive (2000) reanalyze a subset of
the NACRE compilation in a simple, but uniform way in order to
establish a more rigorous error assignment.  Based on these most
recent analyses, and the accuracy with which the WMAP satellite was
able to determine the baryon density, it is clear that a rigorous and
self-consistent prescription for dealing with nuclear data and
deriving accurate representations and uncertainties must be
established.  It is with this main goal in mind, that we have
developed the prescription in the previous section.

There are two kinds of reactions, those induced by neutrons and those
induced by charged particles.  The cross sections for these reactions
are generally decomposed into forms that behave more smoothly than the
cross sections.  In general, low energy cross sections scale with the
square of the de Broglie wavelength, $\sigma \propto \lambda^2 \propto
1/v^2$, where $v$ is the relative velocity between the incident and
target particles.  There are further modifications to this behavior
depending on the type of interactions involve.  The neutron induced
reactions feel only the strong nuclear force. The transmission
probability of a neutron hitting this sharp potential surface is
proportional to $v$, thus the neutron induced reactions can be written
as follows:
\beq
\label{eqn:n-ind}
\sigma(E) = \frac{R(E)}{N_Av(E)}
\eeq
where $R(E)$ is usually a smoothly varying function of center of mass
energy, $E$, and constant at low energies.  $N_A$ is Avogadro's
number.  The charge induced reactions feel the long range
electromagnetic force, with a transmission probability exponentially
suppressed by the Sommerfeld parameter, $\zeta$.  The charge induced
cross sections can be decomposed into
\beq
\label{eqn:q-ind}
\sigma(E) = \frac{S(E)\exp{(-2\pi\zeta)}}{E},
\eeq
where $S(E)$ is the astrophysical $S$-factor, and the
Sommerfeld parameter is defined by
\beq
\zeta = Z_1Z_2\alpha \left(\frac{\mu c^2}{2E}\right)^{1/2} = \frac{1}{2\pi}\left( \frac{E_g}{E}\right)^{1/2}.
\eeq
Here the $Z_i$'s and $\mu$ are the charge numbers and reduced mass of
the reactants, $\alpha$ is the fine-structure constant and $E_g =
2\pi^2Z_1^2Z_2^2 \alpha^2\mu c^2$ is the Gamow energy.  The
$S$-factor, $S(E)$, can also be a slowly varying function of energy.

In the following we evaluate best-fits and uncertainties in $R(E)$ and
$S(E)$, following our above statistical procedure.  We use polynomial
fitting functions, $y(x) = \sum_{n=0}^Na_nx^n$, where the degree N of
the polynomial is allowed vary until a minimum $\chi^2_{\nu}$ is
found.  The data used in the following discussion has been gathered
largely with the use of the NNDC's website \cite{nndc}.


\subsubsection{$n$-decay and Newton's G${}_N$}
\label{sect:2.3.1}

The lifetime of the neutron and Newton's G${}_N$ are key in
determining the amount of \he4, being dependent on the neutron
abundance at the deuterium bottleneck, they also dominate the \he4
uncertainty.  The lifetime of the neutron is key in determining the
rate of neutron-proton inter-conversion.  Reactions such as
$n+\nu_e\rightarrow p^{+}+e^{-}$ and $n\rightarrow
p^{+}+e^{-}+\bar{\nu}_e$, have a common normalization and thus can be
scaled with the mean neutron lifetime.  The propagation of the neutron
lifetime uncertainty into the light element abundance predictions was
first explored by Olive {\em et al}.~\cite{ostys} and in subsequent
works~\cite{kr90,skm,kk95,hata,flsv,nb00,cfo1,cvcr}.  Newton's
G$_N$ enters into the BBN calculation through the universal expansion
physics.  The effect of the gravitational constant's uncertainty has
been previously examined by Scherrer~\cite{scherrer}, and agrees well
with this work's results.  We adopt the recommended neutron lifetime
and Newton's G${}_N$ from the Particle Data Group (2002) \cite{pdg}
with $\tau_n = 885.7 \pm 0.8$ sec and G${}_N = (6.673\pm 0.010)\times
10^{-8}$ (cgs).


\subsubsection{$p(n,\gamma)d$}
\label{sect:2.3.2}

\begin{figure}[ht]
\begin{center}
\psfig{file=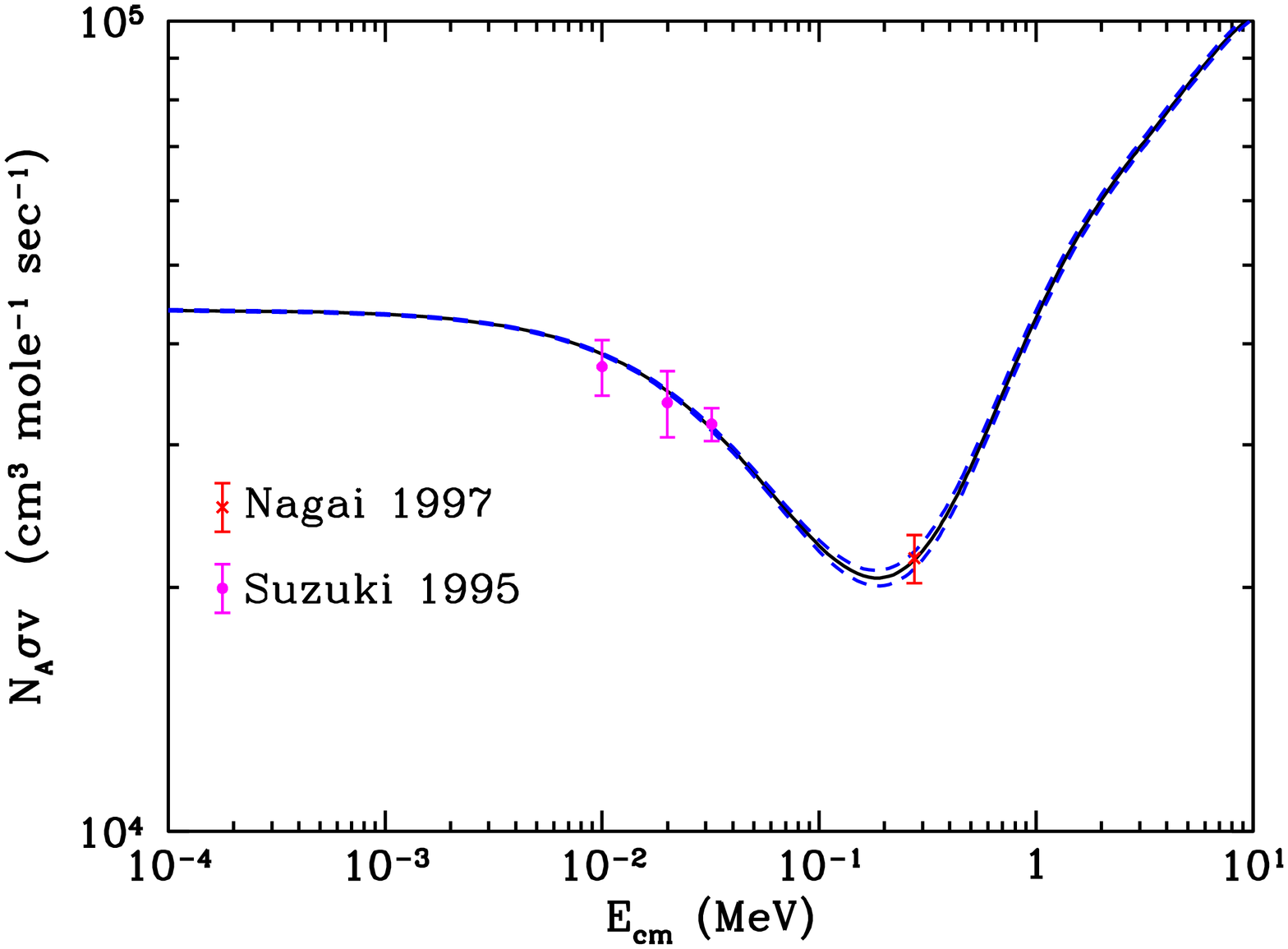,width=4.5in,angle=270}
\caption{The reaction rate data for $p(n,\gamma)d$.  The solid line
represents the best fit, whilst the dashed the 1-sigma error bars.
The fit is an R-matrix calculation by Hale \& Johnson (2003) \cite{hale}.
The data is shown with their respective 1-sigma error bars.}
\label{fig:2-npg}
\end{center}
\end{figure}

Knowing the $p(n,\gamma)d$ reaction is key in determining the end of
the deuterium bottleneck and thus the onset of big bang
nucleosynthesis.  This radiative capture reaction is measured sparsely
in the energy range of interest for BBN, $.01-1.0$ MeV.  It is because
of this lack of data that we must rely on a constrained R-matrix fit
using elastic $p-p$, $n-p$ scattering, and both unpolarized and
polarized $\gamma-d$ photo-disintegration data, in addition to the
sparse $np$-capture data of Nagai 1997~\cite{nagai} and Suzuki
1995~\cite{suzuki}.  We adopt the R-matrix calculation of Hale and
Johnson (2003) \cite{hale}, who have used the data discussed above to
determine the $np$-capture cross section and its energy dependent
uncertainties.  This information was graciously provided by G. Hale
upon private communication.  This rate is now know to better than
2.5\%, about a factor of 2 improvement over previous studies.  We do
not calculate a systematic error for this reaction.


\subsubsection{$d(p,\gamma)\he3$}
\label{sect:2.3.3}

\begin{figure}[ht]
\begin{center}
\psfig{file=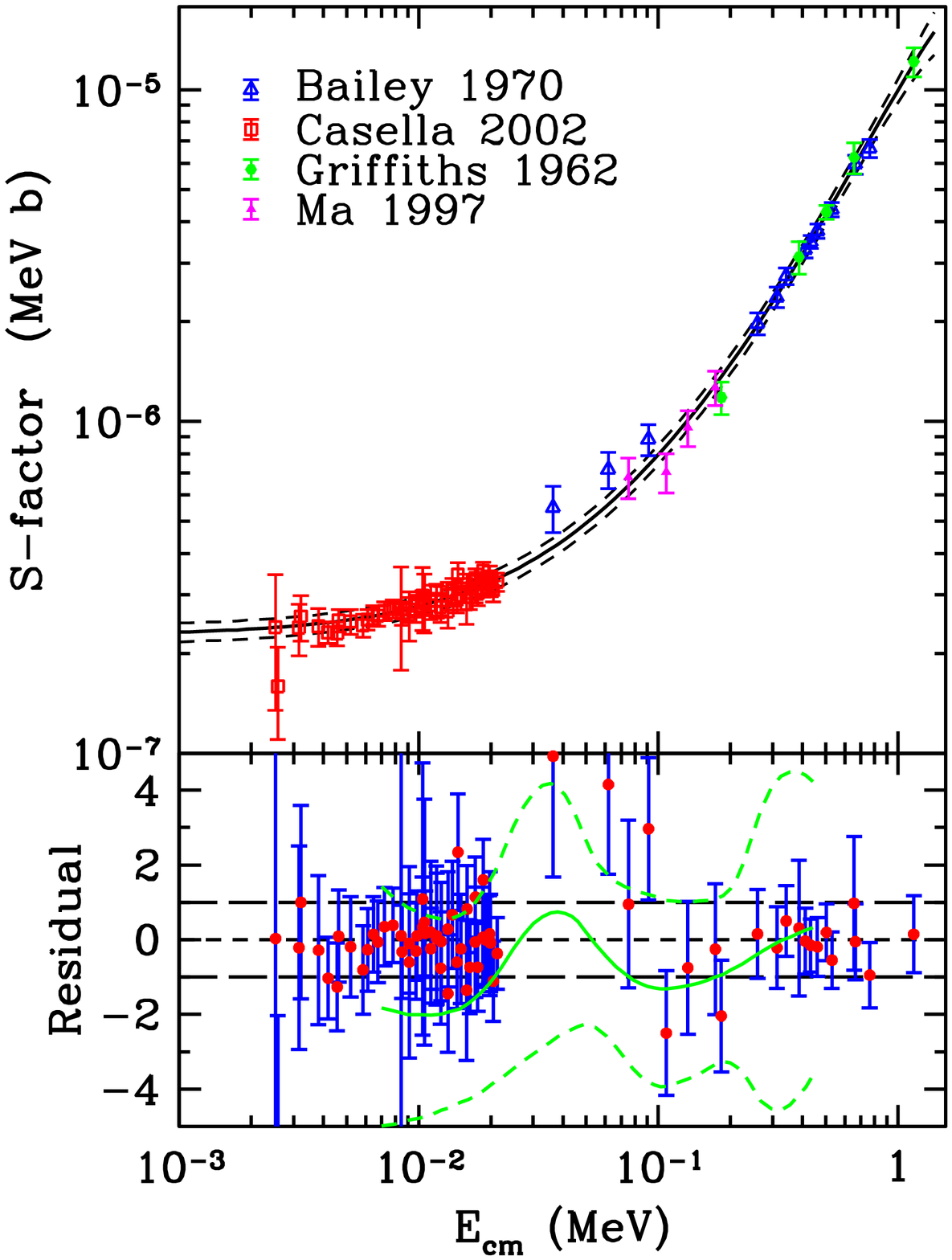,height=4.0in}
\caption{The reaction rate data for $d(p,\gamma)\he3$.  The solid line
represents the best fit, whilst the dashed the 1-sigma error bars.
The data is shown with their respective 1-sigma error bars.  The
bottom panel shows the residual scattering in the data about our best
fit, where our errors are set to $\pm 1$.  The light curves are the
Nollett and Burles \cite{nb00} best fit and $1\sigma$ errors.}
\label{fig:2-dpg}
\end{center}
\end{figure}

The $d(p,\gamma)\he3$ reaction is the first in a chain of reactions
that rapidly burn deuterium after the deuterium bottleneck into \he3
and eventually \he4.  There are few data sets for this reaction in the
BBN energy range.  We consider the data sets of
Bailey (1970) \cite{dpgbailey}, Griffiths (1962,1963)
\cite{dpggriffiths}, Ma (1997) \cite{dpgma},
Schmid (1995,1996) \cite{dpgschmid} and Casella (2002)
\cite{dpgcasella}.  Some of these data sets warrant detailed
consideration.  The Casella data is the most recent measurement of
this cross section and serves to anchor the low energy behavior of
this reaction.  This data has not been included in older analyses,
only in this and two more recent BBN compilations by Cuoco {\em et
al}~\cite{cimm} and Coc {\em et al}.~\cite{cvdaa}.  It has been
suggested that the 1963 Griffiths and 1970 Bailey experiments used
incorrect stopping powers, and thus their low energy behavior is $\sim
15$ \% too high \cite{dpgma,dpgschmid}. Since the Casella data
dominates the low energy behavior of the cross section, inclusion of
the Bailey data does not affect this region of the cross section, thus
we find no reason to omit it from our analysis.  The 1963 Griffiths
data however, does not have a clear discussion of the normalization
uncertainties, thus we exclude this data set from our analysis. The
Schmid data sets suffer from poor energy resolution, with typical
uncertainties in energy greater than 10\%, which have not been
included in their cross section errors.  We thus exclude the Schmid
data sets from the analysis. The inclusion of the new Casella data,
greatly increases the accuracy of the cross section, when compared to
previous analyses.

The discrepancy systematic error is $\delta_{disc} = 0.0345$, the
intrinsic normalization error is $\delta_{norm} = 0.0528$ and the
total systematic error is $\delta = 0.0631$. 


\subsubsection{$d(d,n)\he3$}
\label{sect:2.3.4}

\begin{figure}[ht]
\begin{center}
\psfig{file=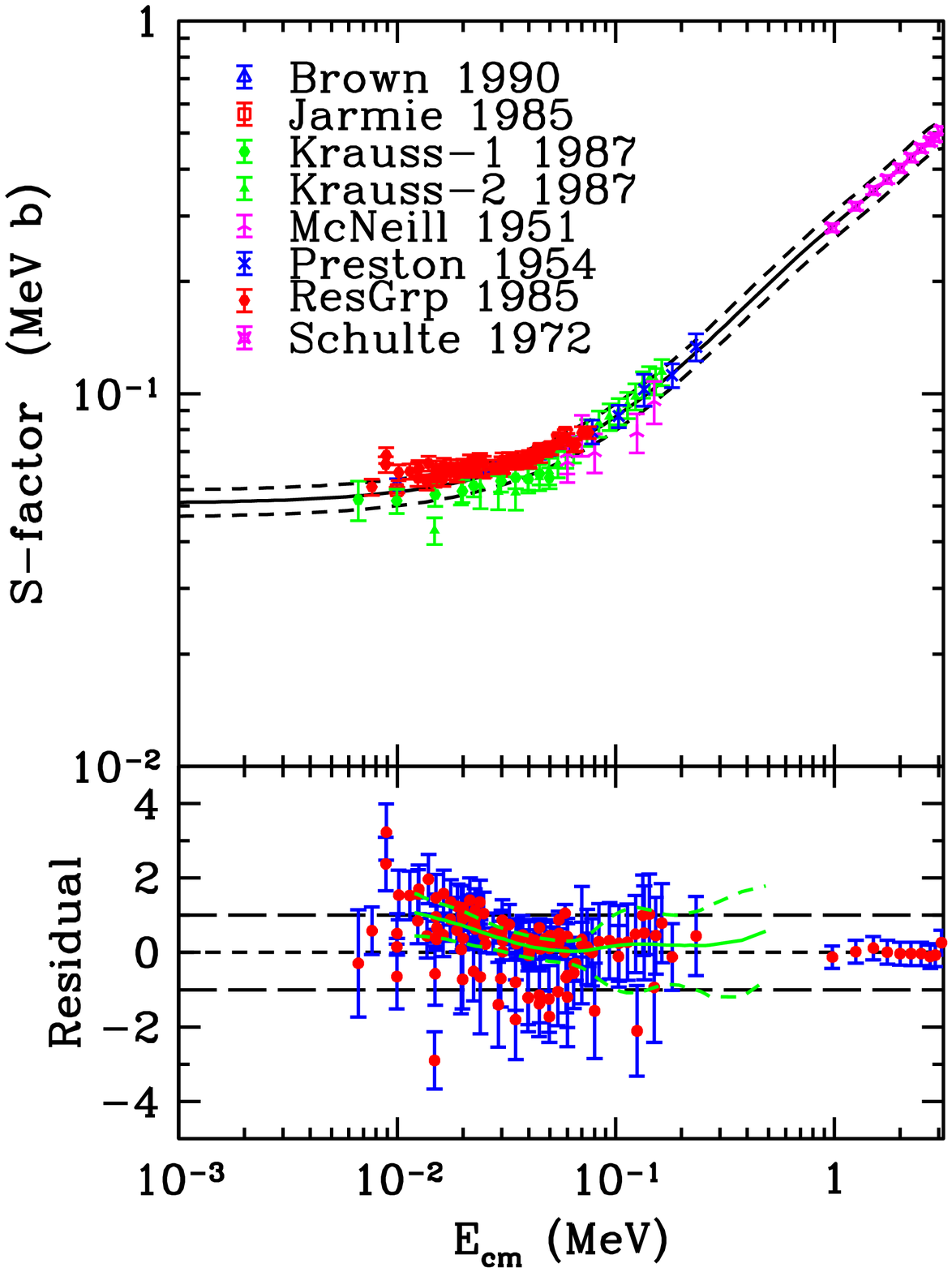,height=4.0in}
\caption{Same as fig.~\ref{fig:2-dpg} but for $d(d,n)\he3$.}
\label{fig:2-ddn}
\end{center}
\end{figure}

The $d(d,n)\he3$ reaction is the dominant deuterium sink during
primordial nucleosynthesis.  We consider the data sets of Brown (1990)
\cite{brown}, Krauss (1987) \cite{krauss}, Ganeev (1958)
\cite{ganeev}, Arnold (1954) \cite{arnold}, McNeill (1951)
\cite{mcneill}, Research Group (1985) \cite{resgrp}, Preston
(1954) \cite{preston}, Jarmie (1985) \cite{ddn-jarmie} and Schulte
(1972) \cite{ddn-schulte}.  Of these data sets, inconsistancies in the
Ganeev data set found on the NNDC website~\cite{nndc}, create
difficulties when trying to separate systematic errors from the total
errors presented (e.g. unphysical statistical errors), thus we exclude
this data set.  The Arnold data exists only as a smoothed data set.
This smoothing will artificially increase this data set's weight on
the fit, thus we exclude this data set.  The high energy Schulte data
helps smoothly interpolate the gap between it and the low energy data.
One may notice that the fitted curve falls below a majority of the
data, a seemingly bad ``chi-by-eye'' fit.  The eye is misleading in
this case.  Since we are treating the correlations between data points
explicitly, it is important to understand its impact.  As we
determined in the previous section, the statistical uncertainty plays
a larger role than the total uncertainty of a particular data point.
When a data set has very small statistical uncertainties, it gets more
weight when determining the fit.  This is exactly what we are seeing
here.  Though the Brown and Research Group data have small
normalization errors, their statistical errors are large when compared
to the statistical uncertainties in the Krauss data, where $\sigma \la
1$ \%.  Even though the Krauss data has larger normalization
uncertainties, its statistical errors are significantly smaller than
other data sets, and thus the Krauss data dominates the low energy
behavior of the fit.  It is interesting to note that if we turn off
the correlations between data points and adopt the total uncertainties
as the representative errors, we reproduce the mean value of the
Nollett and Burles curve.

The discrepancy systematic error is $\delta_{disc} = 0.0369$, the
intrinsic normalization error is $\delta_{norm} = 0.0400$ and the
total systematic error is $\delta = 0.0544$.


\subsubsection{$d(d,p)t$}
\label{sect:2.3.5}

\begin{figure}[ht]
\begin{center}
\psfig{file=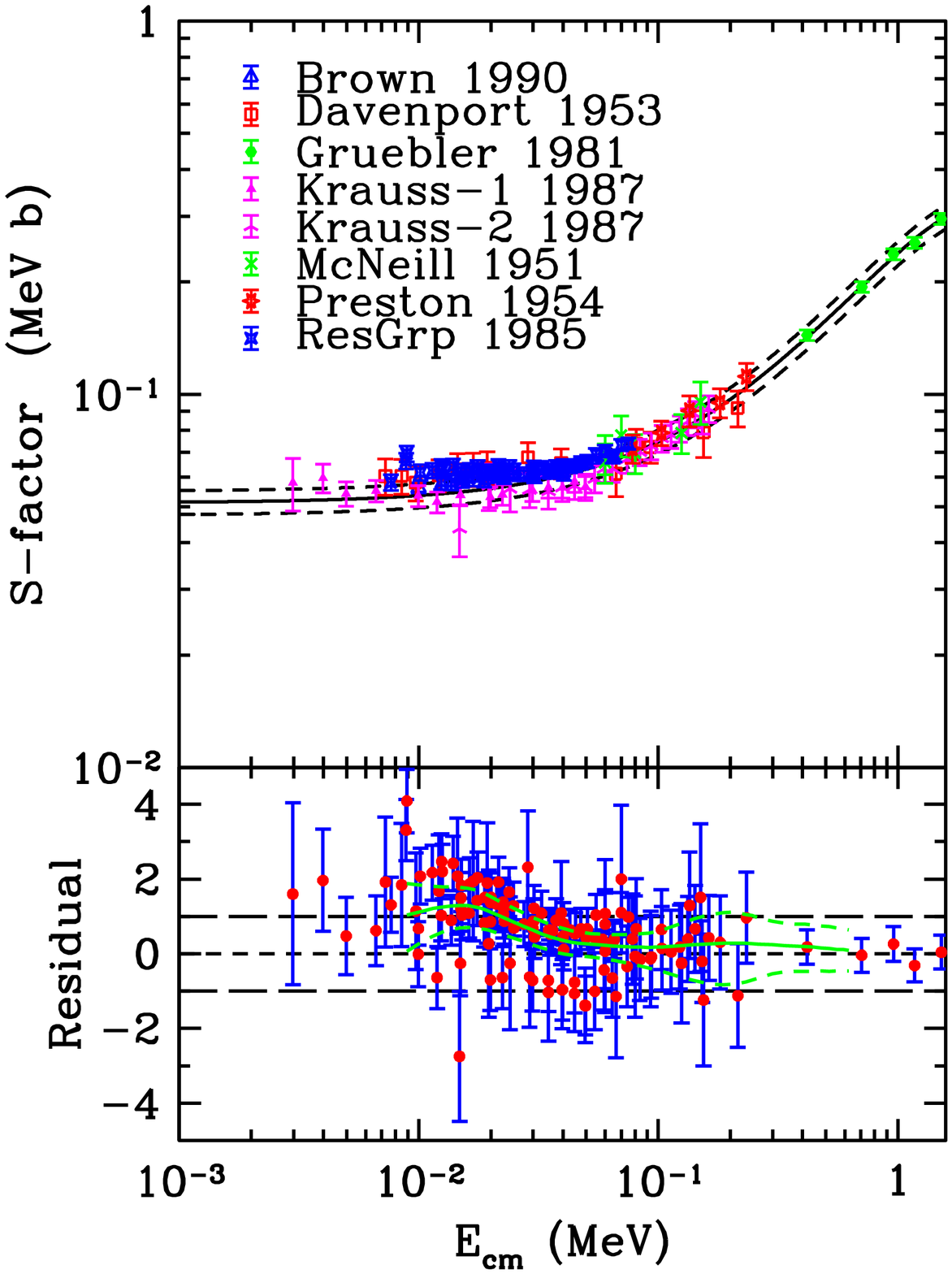,height=4.0in}
\caption{Same as fig.~\ref{fig:2-dpg} but for $d(d,p)t$.}
\label{fig:2-ddp}
\end{center}
\end{figure}

The $d(d,p)t$ reaction is very similar to its mirror $d(d,n)\he3$
reaction, both in shape and magnitude.  We consider the data sets of
Krauss (1987) \cite{krauss},  Brown (1990) \cite{brown}, Preston (1954)
\cite{preston}, Arnold (1954) \cite{arnold}, Davenport (1953)
\cite{ddp-davenport}, Research Group (1985) \cite{resgrp}, Ganeev
(1958) \cite{ganeev}, McNeill (1951) \cite{mcneill}, and
Gruebler (1981) \cite{ddp-gruebler}.  We exclude the data sets of
Ganeev and Arnold for the same reasons as for the $d(d,n)\he3$
reaction.  We again see the statistical uncertainties in the Krauss
data pulling the fit below the Research Group and Brown data.  Again,
this is entirely due our explicit treatment of the correlations in the
data.  If we turn off the correlations and adopt the total uncertainty
as the representative uncertainty, we again reproduce the mean
value curve of Nollett and Burles.

The discrepancy systematic error is $\delta_{disc} = 0.0487$, the
intrinsic normalization error is $\delta_{norm} = 0.0560$ and the
total systematic error is $\delta = 0.0742$.


\subsubsection{$\he3(n,p)t$}
\label{sect:2.3.6}

\begin{figure}[ht]
\begin{center}
\psfig{file=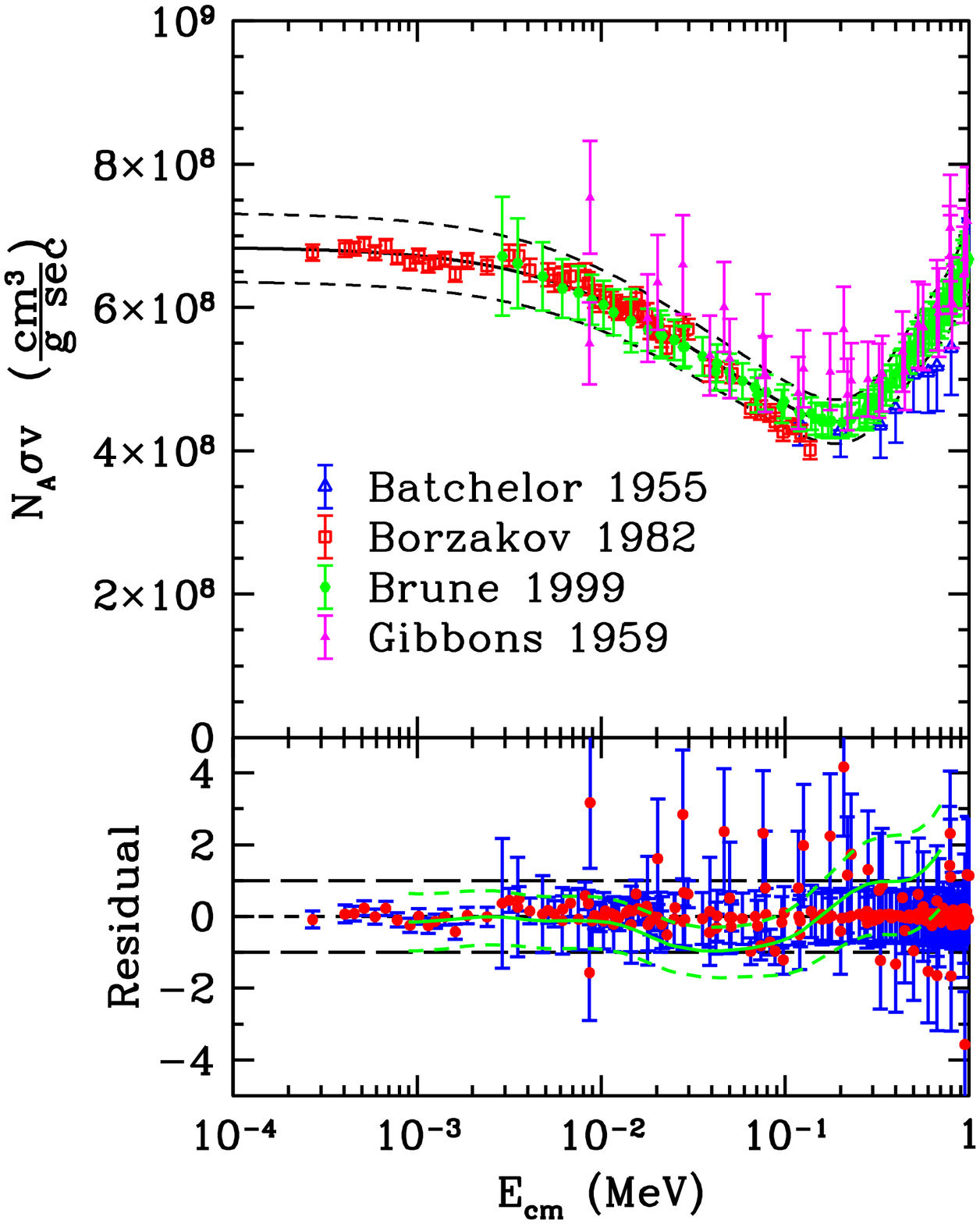,height=4.0in}
\caption{Same as fig.~\ref{fig:2-dpg} but for $\he3(n,p)t$.}
\label{fig:2-3np}
\end{center}
\end{figure}

The $\he3(n,p)t$ reaction is responsible for the inter-conversion of
mass 3 elements, maintaining an equilibrium relation between the two
elements while this rate is fast when compared to the Hubble expansion
rate.  We consider the data sets of Brune (1999) \cite{3np-brune},
Costello (1970) \cite{3np-costello}, Coon (1950) \cite{3np-coon},
Gibbons (1959) \cite{gibbons}, Macklin (1965) \cite{3np-macklin},
Batchelor (1955) \cite{3np-batchelor}, Borzakov (1982)
\cite{3np-borzakov} and Alfimenkov (1980) \cite{3np-alfimenkov}.  We
exclude the Costello data because of poor energy resolution, the Coon
and Macklin data because of little or no error information, and the
Alfimenkov data because the reference was not available.  There is a
lot of data above 1 MeV for this reaction.  In order to fit all of
the data, we would need many fitting parameters.  Since the energy
range relevant for BBN is below 1 MeV, we do not use data above 1
MeV.  As one can see, the fit is dominated by the Brune data.

The discrepancy systematic error is $\delta_{disc} = 0.00703$, the
intrinsic normalization error is $\delta_{norm} = 0.0468$ and the
total systematic error is $\delta = 0.0473$.


\subsubsection{$t(d,n)\he4$}
\label{sect:2.3.7}

\begin{figure}[ht]
\begin{center}
\psfig{file=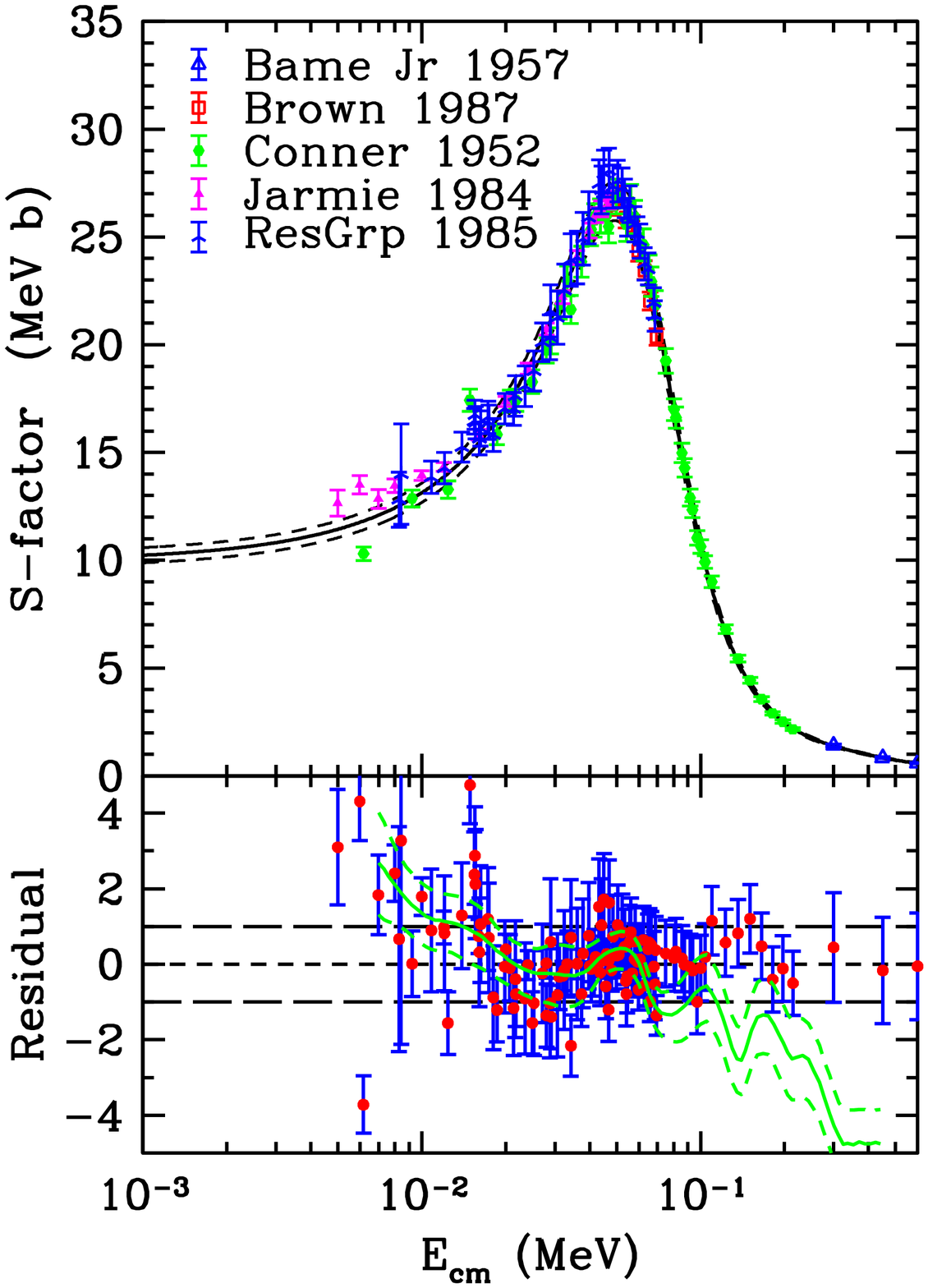,height=4.0in}
\caption{Same as fig.~\ref{fig:2-dpg} but for $t(d,n)\he4$.}
\label{fig:2-tdn}
\end{center}
\end{figure}

The $t(d,n)\he4$ reaction is a main production route to \he4.  We
will consider the data sets of Allan (1951) \cite{tdn-allan}, Argo
(1952) \cite{tdn-argo}, Arnold (1954) \cite{arnold}, Bame
Jr. (1957) \cite{tdn-bamejr}, Brown (1987) \cite{tdn-brown}, Conner
(1952) \cite{tdn-conner}, Davidenko (1957) \cite{tdn-davidenko},
Jarmie (1984) \cite{tdn-jarmie} and Research Group (1985)
\cite{resgrp}.  We exclude the Allan and Argo data sets because of
 uncertain normalization error assignments, and the Davidenko data
set because the reference was not available.  We again exclude the
Arnold data set because of their smoothing their data.  The Conner
data assumes the cross section is isotropic.  This assumption is good
up to energies of $E\sim 240$ keV, thus we exclude any Conner data
that lie beyond this energy.

The discrepancy systematic error is $\delta_{disc} = 0.0218$, the
intrinsic normalization error is $\delta_{norm} = 0.0401$ and the
total systematic error is $\delta = 0.0456$.


\subsubsection{$\he3(d,p)\he4$}
\label{sect:2.3.8}

\begin{figure}[ht]
\begin{center}
\psfig{file=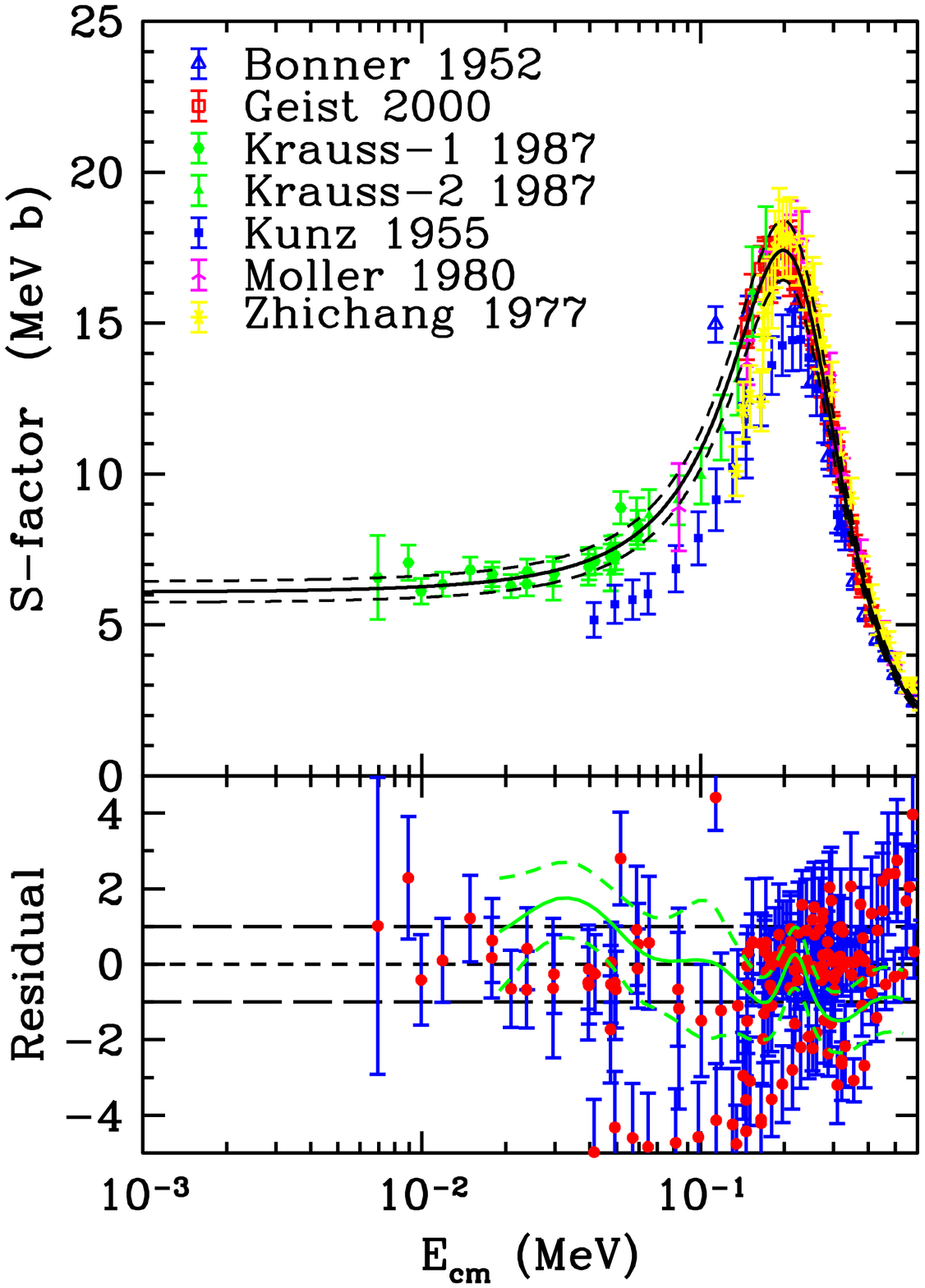,height=4.0in}
\caption{Same as fig.~\ref{fig:2-dpg} but for $\he3(d,p)\he4$.}
\label{fig:2-3dp}
\end{center}
\end{figure}

The $\he3(d,p)\he4$ reaction is also a main route for producing \he4.
We consider the data sets of Arnold (1954) \cite{arnold}, Bonner
(1952) \cite{3dp-bonner}, Geist (2000) \cite{3dp-geist}, Krauss (1987)
\cite{krauss}, Kunz (1955) \cite{3dp-kunz}, Moller (1980)
\cite{3dp-moller} and Zhichang (1977) \cite{3dp-zhichang}.  We exclude
the Arnold data again, because of their smoothing the data.

The discrepancy systematic error is $\delta_{disc} = 0.0268$, the
intrinsic normalization error is $\delta_{norm} = 0.0605$ and the
total systematic error is $\delta = 0.0662$.


\subsubsection{$\he3(\alpha,\gamma)\be7$}
\label{sect:2.3.9}

\begin{figure}[ht]
\begin{center}
\psfig{file=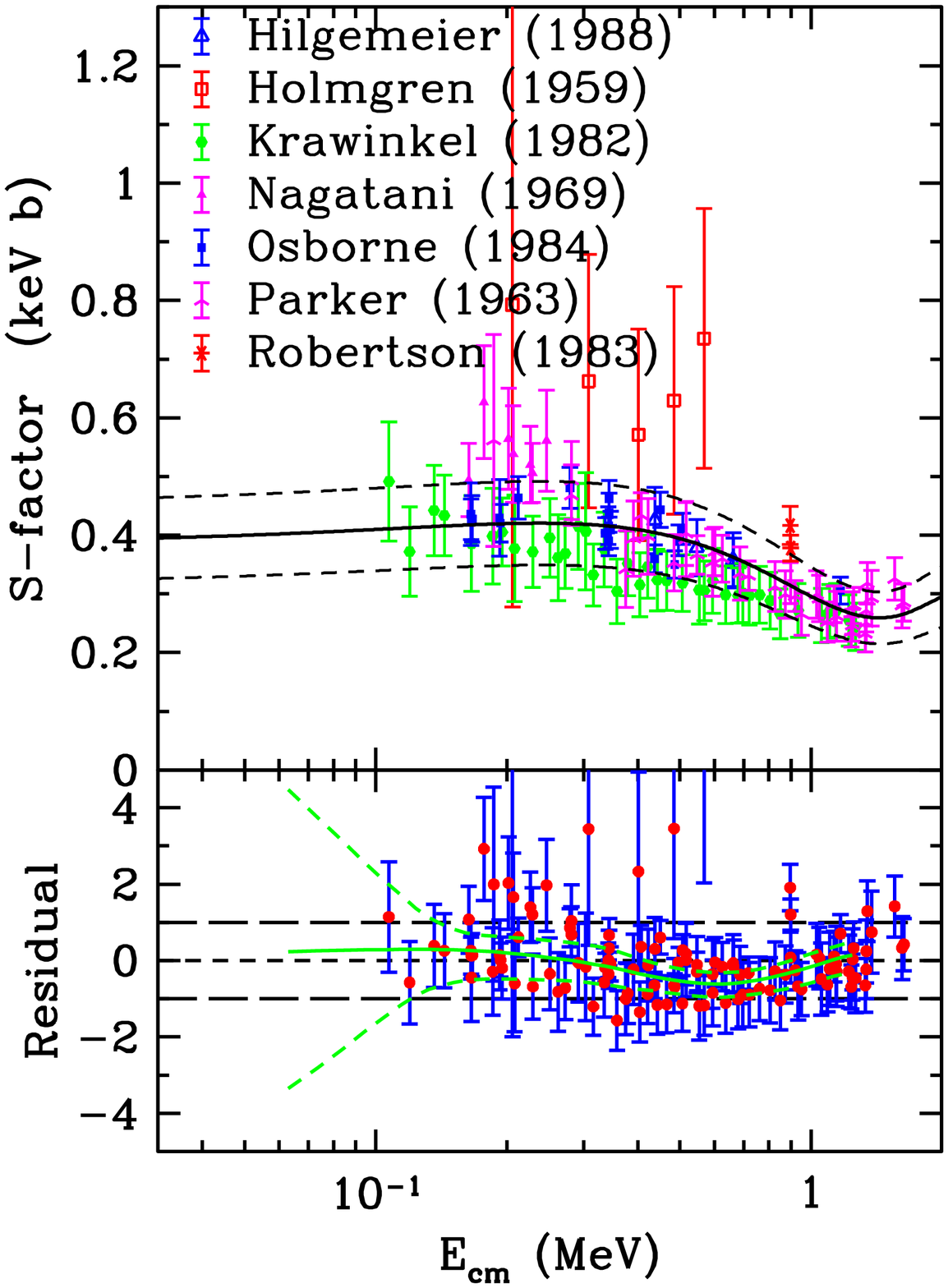,height=4.0in}
\caption{Same as fig.~\ref{fig:2-dpg} but for $\he3(\alpha,\gamma)\be7$.}
\label{fig:2-3ag}
\end{center}
\end{figure}

The $\he3(\alpha,\gamma)\be7$ reaction is responsible for the
production of \li7 in a high baryon density
($\eta\ga3\times10^{-10}$) universe.  Its uncertainty dominates the
prediction of the \li7 abundance prediction.  We consider the data
sets of Holmgren (1959) \cite{holmgren}, Parker (1963)
\cite{3ag-parker}, Nagatani (1969) \cite{3ag-nagatani}, Krawinkel
(1982) \cite{3ag-krawinkel}, Robertson (1983)
\cite{3ag-robertson}, Hilgemeier (1988) \cite{3ag-hilgemeier} and
Osborne (1984) \cite{3ag-osborne}.  Following the suggestion of
Hilgemeier, we renormalize the Krawinkel data by the factor 1.4,
correcting the helium gas density.

The discrepancy systematic error is $\delta_{disc} = 0.1482$, the
intrinsic normalization error is $\delta_{norm} = 0.0814$ and the
total systematic error is $\delta = 0.1691$.

This reaction is also very important for stellar physics, in
particular neutrino production.  The low energy behavior of this
reaction rate determines the flux of \be7 and \b8 neutrinos coming
from the Sun.  We believe it is inappropriate to base the low energy
value on an average of extrapolated points, and recommend our adopted
method of a global analysis of the data and its uncertainties and then
extrapolating a low energy value.  We get a value of $S_{34}(0) = 
 (1.0 \pm 0.169)(0.386 \pm 0.020) = 0.386 \pm 0.068$ keV b for the
astrophysical $S$-factor.  This is 
significantly lower than the values determined by Adelberger {\em et
al}. $S_{34}^{\rm Adlb} = 0.53 \pm 0.05$ keV b \cite{Adelberger},
the NACRE collaboration $S_{34}^{\rm nacre} = 0.54 \pm 0.09$ keV b
\cite{nacre} and the Cyburt, Fields and Olive \cite{cfo1} renormalized 
NACRE rate $S_{34}^{\rm CFO} = 0.50 \pm 0.05$ keV b, though all
determinations are based primarily on the same data.  The Osborne data
dominates the fit at low energy, causing the downward turn of the
$S$-factor.  This turn is also seen in the Nollett and Burles
compilation~\cite{nb00}.  The model independent approach adopted in
this work and in the work of Nollett and Burles should not be used for
extrapolation, as these methods are meant to describe the data alone
and thus are only valid where data exists.  However, the inclusion of
any theory fitting this data will still have to include the systematic
errors similar to the ones discussed in this work.  More measurements
with $E\la0.5$ MeV will be able to more precisely determine
$S_{34}(0)$.


\subsubsection{$t(\alpha,\gamma)\li7$}
\label{sect:2.3.10}

\begin{figure}[ht]
\begin{center}
\psfig{file=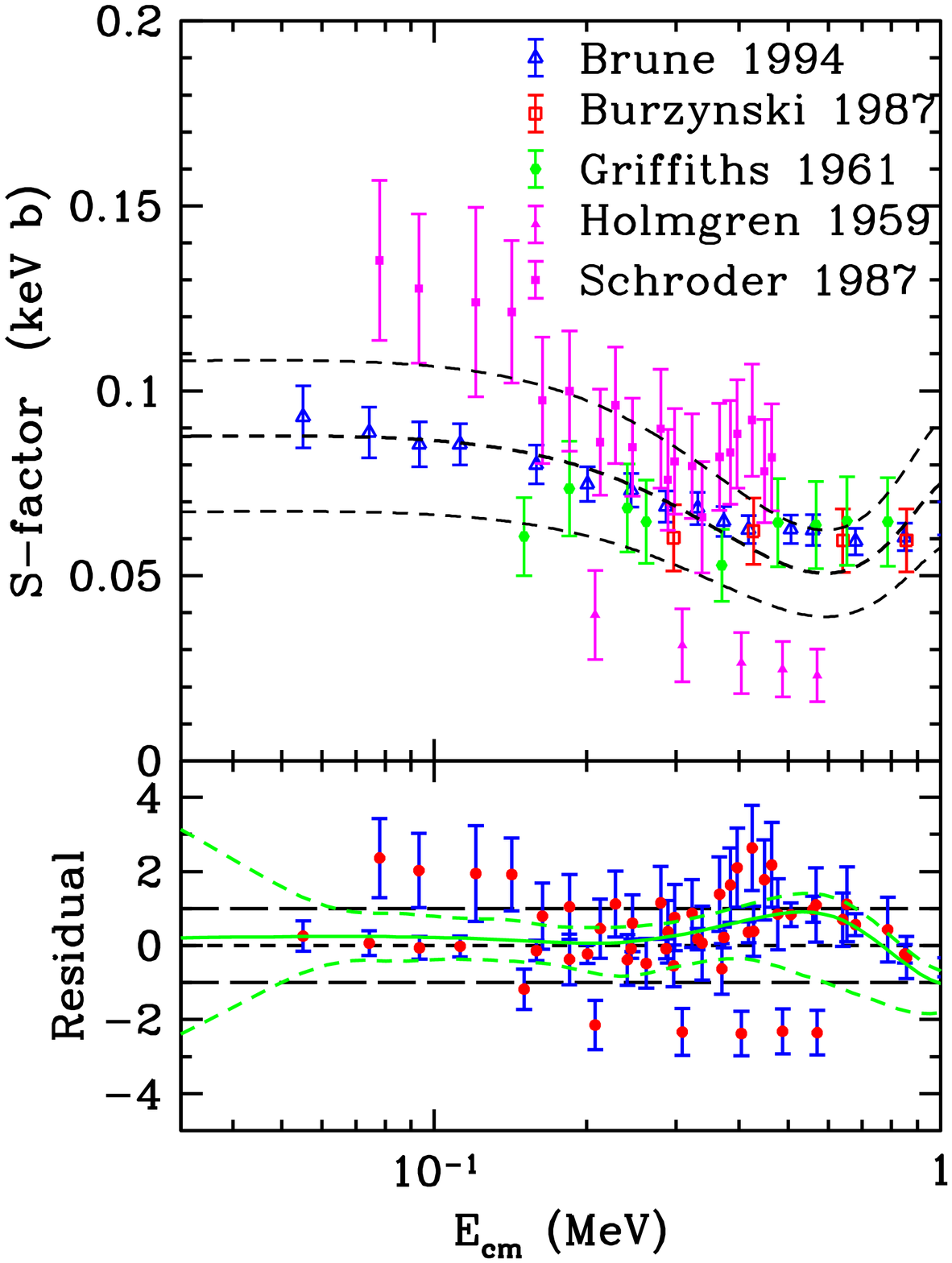,height=4.0in}
\caption{Same as fig.~\ref{fig:2-dpg} but for $t(\alpha,\gamma)\li7$.}
\label{fig:2-tag}
\end{center}\end{figure}

The $t(\alpha,\gamma)\li7$ reaction is important for \li7 production
in a low baryon density ($\eta\la3\times10^{-10}$) universe.  Its
uncertainty dominates the theory prediction of \li7's abundance here.
We consider the data sets of Brune (1994) \cite{tag-brune}, Burzynski
(1987) \cite{tag-burzynski}, Griffiths (1961) \cite{tag-griffiths},
Holmgren (1959)
\cite{holmgren}, Schroder (1987) \cite{tag-schroder} and
Utsunomiya (1990) \cite{tag-utsunomiya}.  We exclude the Utsunomiya
data set because of the lack of a normalization error discussion.  Smith,
Kawano and Malaney \cite{skm} and Nollett and Burles \cite{nb00} make
the point that these Coulomb-breakup measurements are not
yet reliable as this process is not yet completely understood, thus
making the case for new experiments to be performed with $E\la0.2$ MeV.

It is clear that the Holmgren and Schroder data are far from the best
fit curve, outside of their assigned normalization errors.  The
visible discrepancy is forcing the sytematic error to be quite large.
The Holmgren data also pulls the $S$-factor fit down at $E\sim0.6$
MeV.  If reason, other than the visible discrepancy exists to exclude
these data, the fit would be dominated by the high precision Brune
data with an overall 6\% normalization error.

The discrepancy systematic error is $\delta_{disc} = 0.1788$, the
intrinsic normalization error is $\delta_{norm} = 0.1468$ and the
total systematic error is $\delta = 0.2313$.


\subsubsection{$\be7(n,p)\li7$}
\label{sect:2.3.11}

\begin{figure}[ht]
\begin{center}
\psfig{file=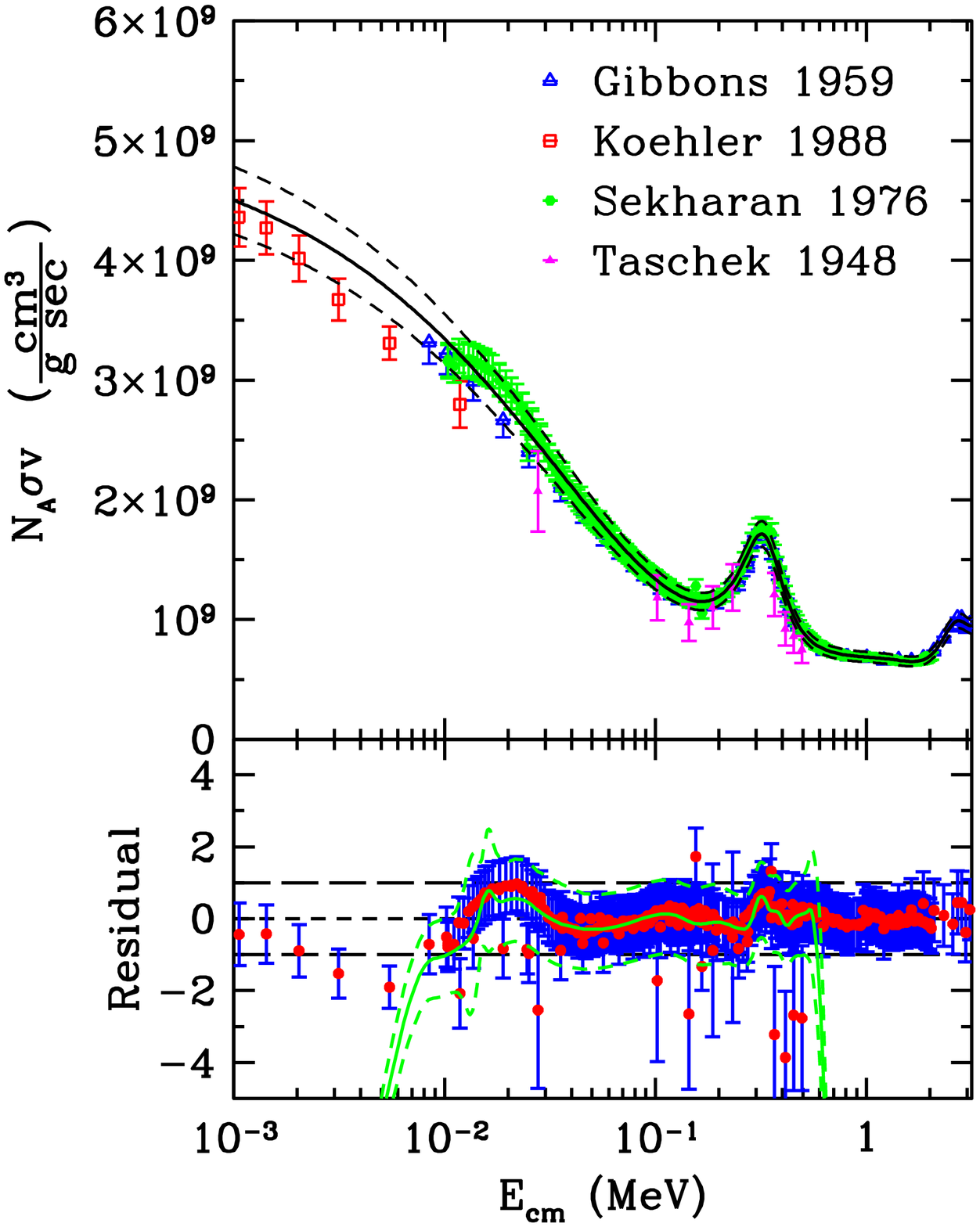,height=4.0in}
\caption{Same as fig.~\ref{fig:2-dpg} but for $\be7(n,p)\li7$.}
\label{fig:2-7np}
\end{center}
\end{figure}

The $\be7(n,p)\li7$ reaction is responsible for the inter-conversion
of mass 7 elements at high baryon density ($\eta\ga3\times10^{-10}$).
This reaction has only one data set in the exoergic direction.  The
data set of Koehler (1988)
\cite{7np-koehler}.  This data set does not extend very far into the
energy range of interest for BBN.  We must rely on the data for the
endoergic reverse reaction, $\li7(p,n)\be7$.  We consider the data
sets of Gibbons (1959) \cite{gibbons}, Sekharan (1976)
\cite{7np-sekharan} and Taschek (1948) \cite{7np-taschek}.  We use the
principal of detailed balance to transform the $\li7(p,n)\be7$ data
into $\be7(n,p)\li7$ data.  Using the $Q$-value from Audi and Wapstra (1995)
\cite{audi} available at the US Nuclear Data Program website
\cite{USNDP}, $Q=1.644168\pm0.000668$ MeV.  We ignore the lowest energy
points derived from the reverse rate as they are sensitive to the
precise value of $Q$, ignoring values that change significantly when
$Q$ is varied within its uncertainties.  We should note that the
Koehler data extends down to well below 1 eV, we choose not to show
the data as its roughly constant and to emphasize the energy range
important for primordial nucleosynthesis.

The discrepancy systematic error is $\delta_{disc} = 0.0159$, the
intrinsic normalization error is $\delta_{norm} = 0.0448$ and the
total systematic error is $\delta = 0.0475$.


\subsubsection{$\li7(p,\alpha)\he4$}
\label{sect:2.3.12}

\begin{figure}[ht]
\begin{center} 
\psfig{file=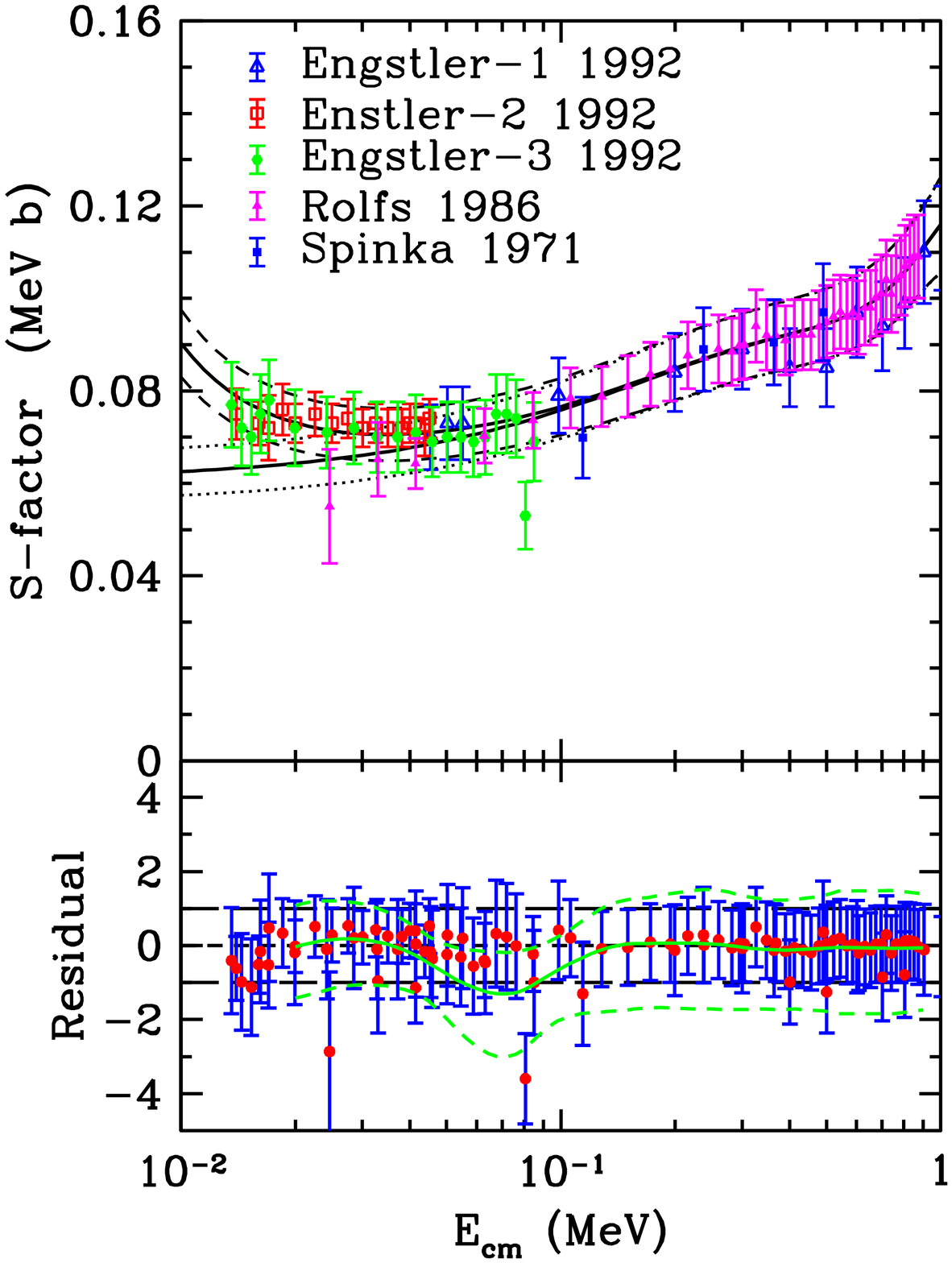,height=4.0in}
\caption{Same as fig.~\ref{fig:2-dpg} but for $\li7(p,\alpha)\he4$.
Also shown is the $e$-screening corrected $S$-factor and its uncertainty.}
\label{fig:2-7pa}
\end{center}
\end{figure}

The $\li7(p,\alpha)\he4$ reaction is the dominant destruction channel
of \li7 at low baryon densities ($\eta\la3\times10^{-10}$).  We
consider the data sets of Engstler (1992) \cite{7pa-engstler}, Harmon
(1989) \cite{7pa-harmon}, Lee (1969) \cite{7pa-lee}, Rolfs (1986)
\cite{7pa-rolfs} and Spinka (1971) \cite{7pa-spinka}.  We exclude the
Harmon data because it is based on a measurement relative to
$\li6(p,\alpha)\he3$ at energies $E\ga150$ keV by Shinozuka {\em et
al} (1979) \cite{7pa-shinozuka}.  All but 3 points lie below this
energy range, thus this measurement relative to $\li6(p,\alpha)\he3$
is not valid at these energies.  One may consider using the 3 points
that are measured at appropriate energies, but it does not change our
fit significantly.  We also exclude the Lee data set as the reference
was unavailable.

This reaction has the largest Gamow energy of all the reactions we
consider, and thus is the most susceptible to electron screening
effects.  In fact, the low energy behavior of this reaction is modified by
electron-screening effects in the experimental set-up.  This behavior
can be parameterized as $\sigma_{\rm exp}(E) = \sigma_{\rm
bare}(E+U_e)$, relating the experimentally measured cross section to
the bare nuclear cross section (i.e. no electron screening), where
$U_e$ is the screening potential \cite{7np-escreening}.  Where $U_e$
is a measure of how much the Coulomb barrier has been reduced due to
electrons screening the bare nucleus (e.g. \li7).  For the experiments
that screening is important (i.e. Engstler), $U_e = 245\pm 45$ eV is found
to be the best fit.  This agrees with the determinations of Engstler
{\em et al}., who find $U_e=300\pm160$ eV using an approximation of
the $e$-screening correction to the observed cross section shown
above.  Englster used the high energy data to determine the best fit,
and then extrapolate this to determine the screening potential.
We have fit all the data, including the electron screening potential
self-consistantly.   

The discrepancy systematic error is $\delta_{disc} = 0.0194$, the
intrinsic normalization error is $\delta_{norm} = 0.0769$ and the
total systematic error is $\delta = 0.0793$.  There is an additional
energy dependent systematic error induced because of our electron screening
correction.  This can be well accounted for by:
\beq
\delta_{e-scr} = 0.02015\left(\frac{\sigma_{U_e}}{45\ {\rm
eV}}\right)\exp{(-15.34E)},
\eeq
again added in quadrature with the other systematics.

\subsection{Thermal Rates}

For the BBN temperature range, Maxwell-Boltzmann phase-space
distributions are an excellent choice for baryons
(eqn.~\ref{eqn:3-lambda}) and the thermal rates become,
\beq
\lambda = N_A\left( \frac{8}{\pi\mu(kT)^3}\right)^{1/2} \int_0^\infty
\sigma(E) E \exp{\left( -\frac{E}{kT}\right)} dE,
\eeq
where $\sigma(E)$ is the cross section, not the standard deviation.
For neutron induced reactions, using eqn.~\ref{eqn:n-ind}, we find
\beq
\lambda = \frac{2}{\sqrt{\pi}(kT)^{3/2}} \int_0^\infty R(E)E^{1/2}
\exp{\left( -\frac{E}{kT}\right)} dE;
\eeq
and for charge induced reactions, using eqn.~\ref{eqn:q-ind}, we find
\beq
\lambda = N_A \left( \frac{8}{\pi\mu(kT)^3}\right)^{1/2} \int_0^\infty
S(E) \exp{\left[ -\frac{E}{kT}- \left(
\frac{E_g}{E}\right)^{1/2}\right]} dE.
\eeq
It is trivial to determine the weighting functions with these
relations.  For the neutron- and charge- induced reactions the
respective weighting functions are:
\beqar
\label{eqn:wet_ni}
W(E,T) &=& \frac{2}{\sqrt{\pi}(kT)^{3/2}}E^{1/2} \exp{\left(
-\frac{E}{kT}\right)}\\ 
\label{eqn:wet_qi}
W(E,T) &=& N_A \left( \frac{8}{\pi\mu(kT)^3}\right)^{1/2} \exp{\left[
-\frac{E}{kT}- \left( \frac{E_g}{E}\right)^{1/2}\right]}.
\eeqar

After these integrals are performed numerically, we must find some
representation of these thermal rates to implement into the BBN code.
We will look at some cases in which the above integrals can be done
analytically.  This will ultimately guide us in determining the
functional forms for these rates.  

Typically, neutron induced reactions follow the $v^{-1}$ behavior
noted in the previous chapter and are particularly smooth over the
data energy range coverage.  Thus a simple polynomial in $E^{1/2}$
will generally suffice.  In this case, the integral can be performed
analytically and the numerical integration serves as a test of the
integrator.

Most of the reactions are non-resonant charge induced reactions.  In
this case the integral cannot be done analytically, particularly at
the temperature ranges relevant for BBN.  In order to understand the
reason for this and gain some insight for a possible functional form,
we will look at the case where the temperature is much smaller than
the Gamow energy, $kT\ll E_g$.  We would like to turn the weighting
function in eqn~\ref{eqn:wet_qi} into something more familiar, like a
gaussian.  To do this, we Taylor expand the argument of the exponent,
about some energy $E_0$, defined such that the first derivative with
respect to energy is zero at $E_0$.  Doing this we find, $E_0/E_g =
(kT/2E_g)^{2/3}$.  Also needed for this analysis is the width of this
gaussian, which we can find by evaluating the second derivative of the
exponent argument at $E_0$.  We find, $\sigma_0/E_g =
2/\sqrt{3}(kT/2E_g)^{5/6}$.  The relevant perturbative parameter is
the ratio of this width to effective energy, $\sigma_0/E_0 =
(32kT/27E_g)^{1/6}$.  In order for this gaussian approximation of the
integrand to converge, the width must be much smaller than the
effective energy. Not only does this demand $kT\ll E_g$, but since the
power is small $(kT/E_g)^{1/6}$ must be small.  To get 10\%
convergence, $kT\la 10^{-6}E_g$, which is well below the range
relevant for BBN. 
However, if we continue, we find that the rate transforms into:
\beq
\label{eqn:qiform}
\lambda = N_A\left( \frac{8}{\mu\pi}\right)^{1/2}
\frac{\sigma_0}{(kT)^{3/2}}\exp{\left[ -\left(
\frac{27E_g}{4kT}\right)^{1/3}\right]}S_{eff}(E_0). 
\eeq
If $S(E)$ is taken to be a polynomial in $E$, then $S_{eff}(E_0)$ is a
polynomial in $(\sigma_0/E_0)^2\propto kT^{1/3}$, where only the even
powers of our perturbative parameter appear due to the symmetry of the
gaussian.  We adopt this form, allowing the order of the polynomial
describing $S_{eff}(E_0)$ to vary as needed until an accurate fit is
reached.  Reactions with broad resonances modify the above form, with
the Breit-Wigner form, with $E=E_0$;
\beq
\lambda = N_A\left( \frac{8}{\mu\pi}\right)^{1/2}
\frac{\sigma_0}{(kT)^{3/2}}\exp{\left[ -\left(
\frac{27E_g}{4kT}\right)^{1/3}\right]}
\frac{S_{eff}(E_0)}{1+((E_0-E_R)/\Gamma_R/2)^2},  
\eeq
where $E_R$ and $\Gamma_R$ are the resonance parameters and
$S_{eff}(E_0)$ is a polynomial in powers of $kT^{1/3}$.

Reactions with narrow resonances are typically the sum of a
non-resonant piece and a Breit-Wigner form.  Since the resonance is
narrow, we can treat the Breit-Wigner form as a delta function.  In
the case of a neutron induced reaction, the resonant part of the rate
becomes:
\beq
\lambda_{res} = \sqrt{\pi}\frac{\Gamma_RE_R^{1/2}}{(kT)^{3/2}}R(E_R)
\exp{(-E_R/kT)}.
\eeq

We wish to reiterate here, that the cross section fits and their
energy dependent uncertainties are numerically integrated.  These
exact results are subsequently cast into a usable form, fit to one of
the forms mentioned above to within 0.1\%.

In thermonuclear rate compilations, such as this one, it is important
to realize that most compilations rely on the same experimental data
to derive their representations.  Because of this, they must generally
agree with each other over the range of validity.  This range is shown
it table~\ref{tab:Trnge}.  The limits are solely based on the maximum
energy of the data used ($kT_{max} \sim E_{max}/3$).   

\begin{table}[ht]
\begin{center}\caption{This table shows to what maximum temperature
each thermal rate is valid for.  This is solely due to the data used in
the analysis.  Also shown are the systematic errors.  } 
\label{tab:Trnge}
\begin{tabular}{lllll}
\hline\hline
Reactions & $T_{max}$ ($10^9$ K) & $\delta_{disc}$ & $\delta_{norm}$ &
$\delta_{tot}$ \\
\hline
$p(n,\gamma)d$            & 100  & N.A. & N.A. & N.A. \\
$d(p,\gamma)\he3$         & 3.9  & 0.0345 & 0.0528 & 0.0631 \\
$d(d,n)\he3$              & 12.5 & 0.0369 & 0.0400 & 0.0544 \\
$d(d,p)t$                 & 5.8  & 0.0487 & 0.0560 & 0.0742 \\
$\he3(n,p)t$              & 3.9  & 0.0071 & 0.0468 & 0.0473 \\
$t(d,n)\he4$              & 2.3  & 0.0218 & 0.0401 & 0.0456 \\
$\he3(d,p)\he4$           & 2.3  & 0.0268 & 0.0605 & 0.0662 \\
$\he3(\alpha,\gamma)\be7$ & 7.8  & 0.1482 & 0.0814 & 0.1691 \\
$t(\alpha,\gamma)\li7$    & 3.9  & 0.1788 & 0.1468 & 0.2313 \\
$\be7(n,p)\li7$           & 11.7 & 0.0159 & 0.0448 & 0.0475 \\
$\li7(p,\alpha)\he4$      & 3.9  & 0.0194 & 0.0769 & 0.0793 \\
\hline\hline
\end{tabular}
\end{center}
\end{table}

\begin{figure}[ht]
\begin{center}
\psfig{file=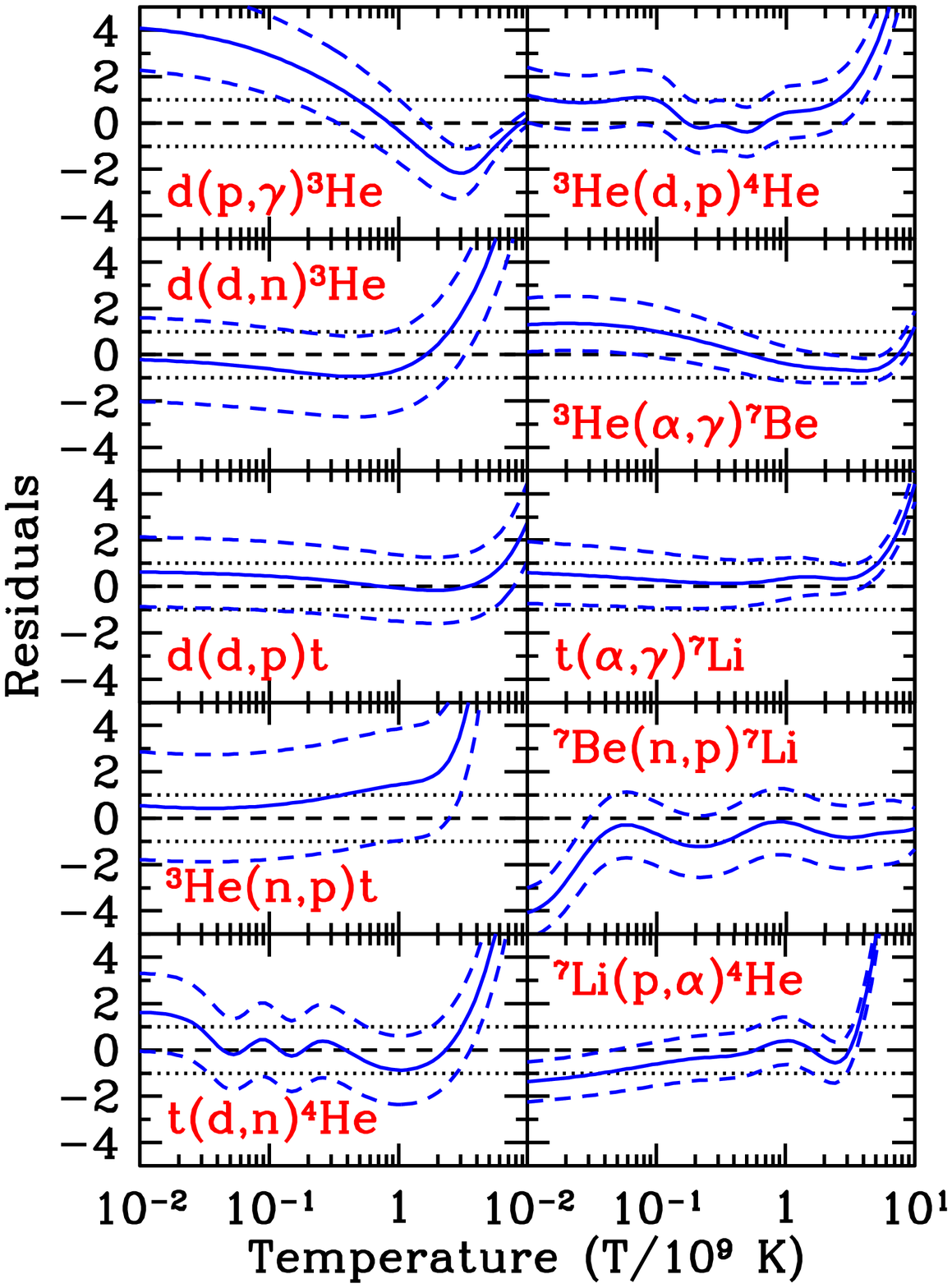,width=4.5in}
\caption{The thermal reaction rate residuals of the Smith, Kawano \&
Malaney (1993) \cite{skm} compilation plotted against temperature in units of
$10^9$ K.  The solid line shows how their mean value compares to this
compilations mean value.  The dashed lines correspond with the
$1\sigma$ errors.}
\label{fig:3-thrmskm}
\end{center}
\end{figure}

We first compare to the BBN reaction rate standard, Smith, Kawano and
Malaney (1993) \cite{skm}.  As we see from fig.~\ref{fig:3-thrmskm},
there is overall agreement with our compilation and theirs.  The
curves tend to diverge at high temperature, where there is no data
pinning down the high energy behavior.  The disagreement with the rate
for $d(p,\gamma)\he3$ is almost entirely due to the use and exclusion
of different data sets.  The disagreement at low temperatures for
$\be7(n,p)\li7$, is most likely due to their taking a minimum energy
when integrating this rate ($E_{min}=1$ keV).  We cannot compare
directly to Nollett and Burles (2000) \cite{nb00} as they do not
present thermal rates, but differences are attributable to the
differences in cross section representations already discussed.  The
``wiggles'' seen in $t(d,n)\he4$, $\he3(d,p)\he4$ and $\be7(n,p)\li7$
are due to the slightly different values adopted for the resonance
parameters of each reaction.

\begin{figure}[ht]
\begin{center}
\psfig{file=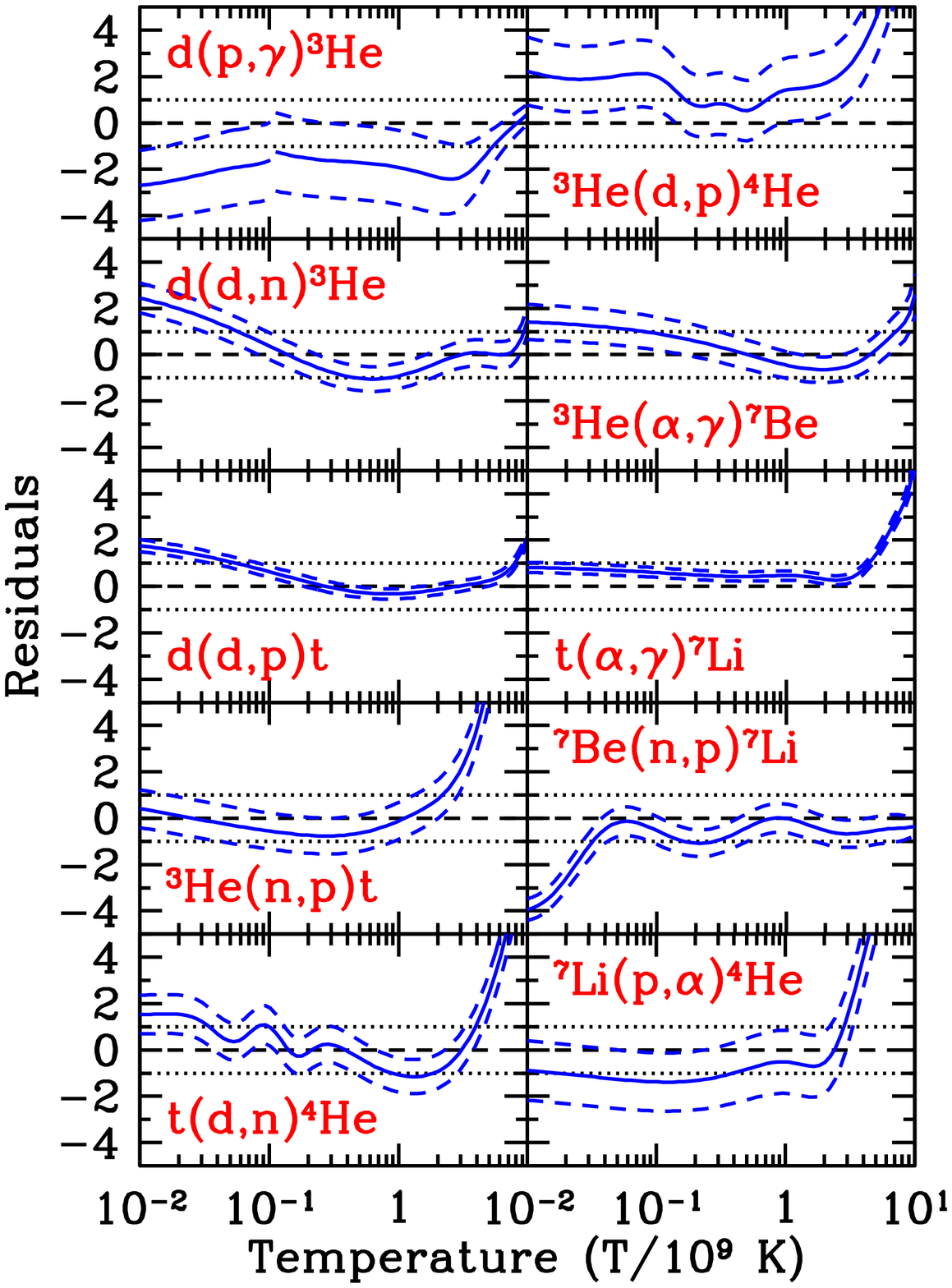,width=4.5in}
\caption{The thermal reaction rate residuals of the Cyburt, Fields \&
Olive  (2001) \cite{cfo1} reanalysis of NACRE \cite{nacre} plotted against
temperature in units of 
$10^9$ K.  The solid line shows how their mean value compares to this
compilations mean value.  The dashed lines correspond with the
$1\sigma$ errors.}
\label{fig:3-thrmcfo}
\end{center}
\end{figure}

We now compare to the work of Cyburt, Fields and Olive (2001)
\cite{cfo1}, which used renormalized NACRE rates and an estimate of 
the errors.  Again, since these compilations are based on most of the
same nuclear data the rates should be similar, as seen in
fig.~\ref{fig:3-thrmcfo}.  An interesting point is that for a majority
of reactions, the Cyburt, Fields and Olive error budget underestimates
the errors, compared to this work.  The intrinsic normalization error
we have included is typically as important as the discrepancy
systematic error, which Cyburt, Fields and Olive assumed to dominate
the error budget.  Again the high temperature portion of the curves
tend to diverge, as there is no data pinning down the high energy
behavior.  We point out that the $d(p,\gamma)\he3$ rate as it is
systematically lower than our rate.  This is entirely due to the NACRE
collaboration's inclusion of data we have excluded.  Also evident are
the differences in the $d(d,n)\he3$ and $d(d,p)t$ rates, where our
compilation falls below the fits adopted by Cyburt, Fields and Olive.
The differences between this compilation and others for these rates,
is that we have included the correlations between data points.  It is
for this reason, that the low energy data that has small statistical,
but large total error is dominating the fits, pulling the low energy
cross section down slightly.

\begin{figure}[ht]
\begin{center}
\psfig{file=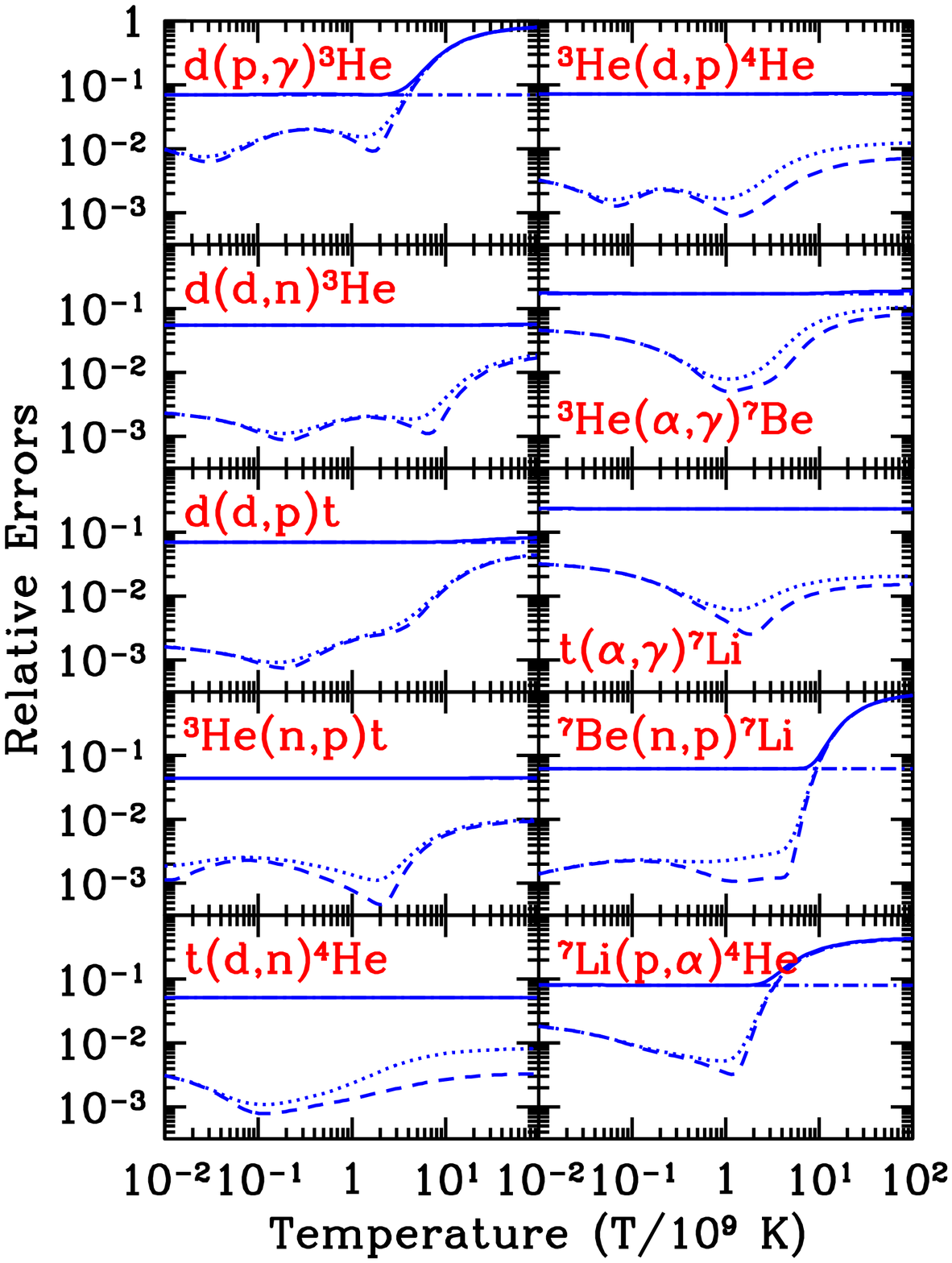,width=4.5in}
\caption{The thermal reaction rate relative errors for the 10
reactions fitted in this compilation.  The dashed and dotted curves
are the thermally averaged statistical errors with and without
treating the energy correlations.  The dashed-dotted curve shows the
total systematic errors, and the solid curve shows the total thermal error.}
\label{fig:3-relerr}
\end{center}
\end{figure}

The overall agreement between different rate compilations is quite
reassuring.  The biggest advantage to our compilation is we have
explicit treatments for dealing with correlated data, and estimating
systematic errors.  These systematic errors dominate over the
statistical uncertainties in all cases, for the temperature range
important for BBN, $T\sim (0.5-1.2) \times 10^9$ K as seen in
fig.~\ref{fig:3-relerr}.  Properly treating the correlation between
fitting parameters when propagating the cross section fits into
thermal rates has a noticeable reduction in the statistical
uncertainties.  This reduction is maximized when there is a cross over
between terms in the fit polynomial, when one term goes from being dominant to
being sub-dominant and vice versa.  It is also reassuring to see the
statistical uncertainty become the dominant contribution to the error
at high temperature or energies, where there is no data.

\subsection{Light Element Predictions}

\begin{figure}[ht]
\begin{center}
\psfig{file=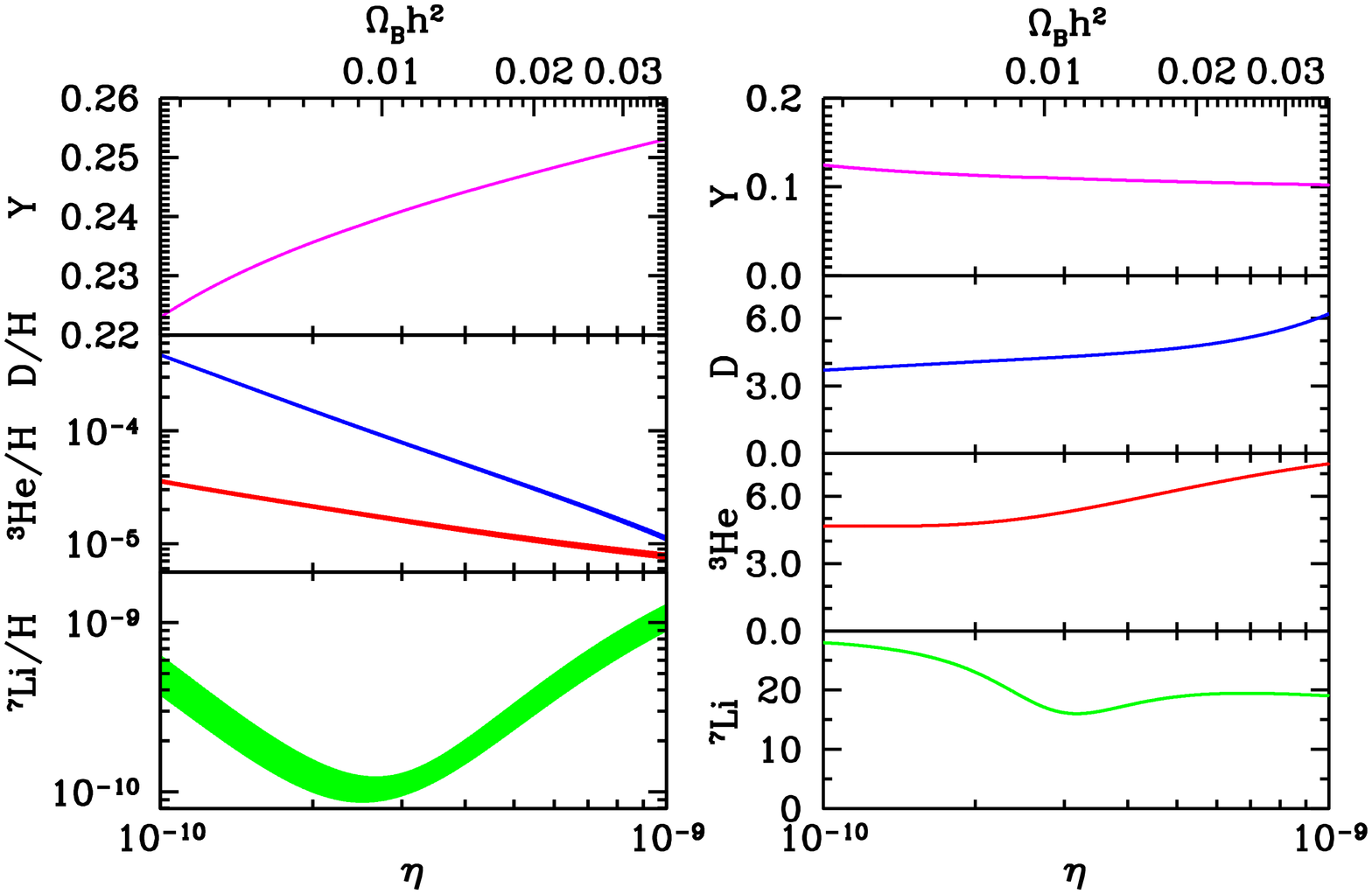,width=4.0in, angle=270}
\caption{Shown in the left panel are the light element predictions
using this works nuclear rate compilation and uncertainties.  The mass
fraction of \he4 ($Y_p$) and the mole fractions, ${\rm D/H}$,
$\he3/{\rm H}$ and $\li7/{\rm H}$
are plotted against the baryon-to-photon ratio.  The width of each curve
represents the 1 $\sigma$ errors in the light element predictions.
The right panel shows the relative uncertainties in percent of the light
element predictions.
\label{fig:4-abvseta}}
\end{center}
\end{figure}

Adopting the thermonuclear reaction rates discussed in the previous
section, we discuss their impact on BBN predictions and on the general
concordance of the BBN predictions with the light element observations
and the CMB.  Shown in figure~\ref{fig:4-abvseta} are the light
element abundance predictions and their percent errors using this
work's nuclear compilation.

Before we examine the concordance between this compilation's
predictions and observations, we should verify the agreement of
previous compilations and qualify their differences.  As has been
discussed in previous chapters, the BBN compilations of Smith, Kawano
and Malaney \cite{skm}, Nollett and Burles \cite{nb00} and Cyburt,
Fields and Olive \cite{cfo1}, should all roughly agree as they are
largely based on the same nuclear data.  Any differences in their
predictions will arise entirely from each compilation's derivation of
reaction rates and their uncertainties, and the data each uses.

\begin{figure}[ht]
\begin{center}
\psfig{file=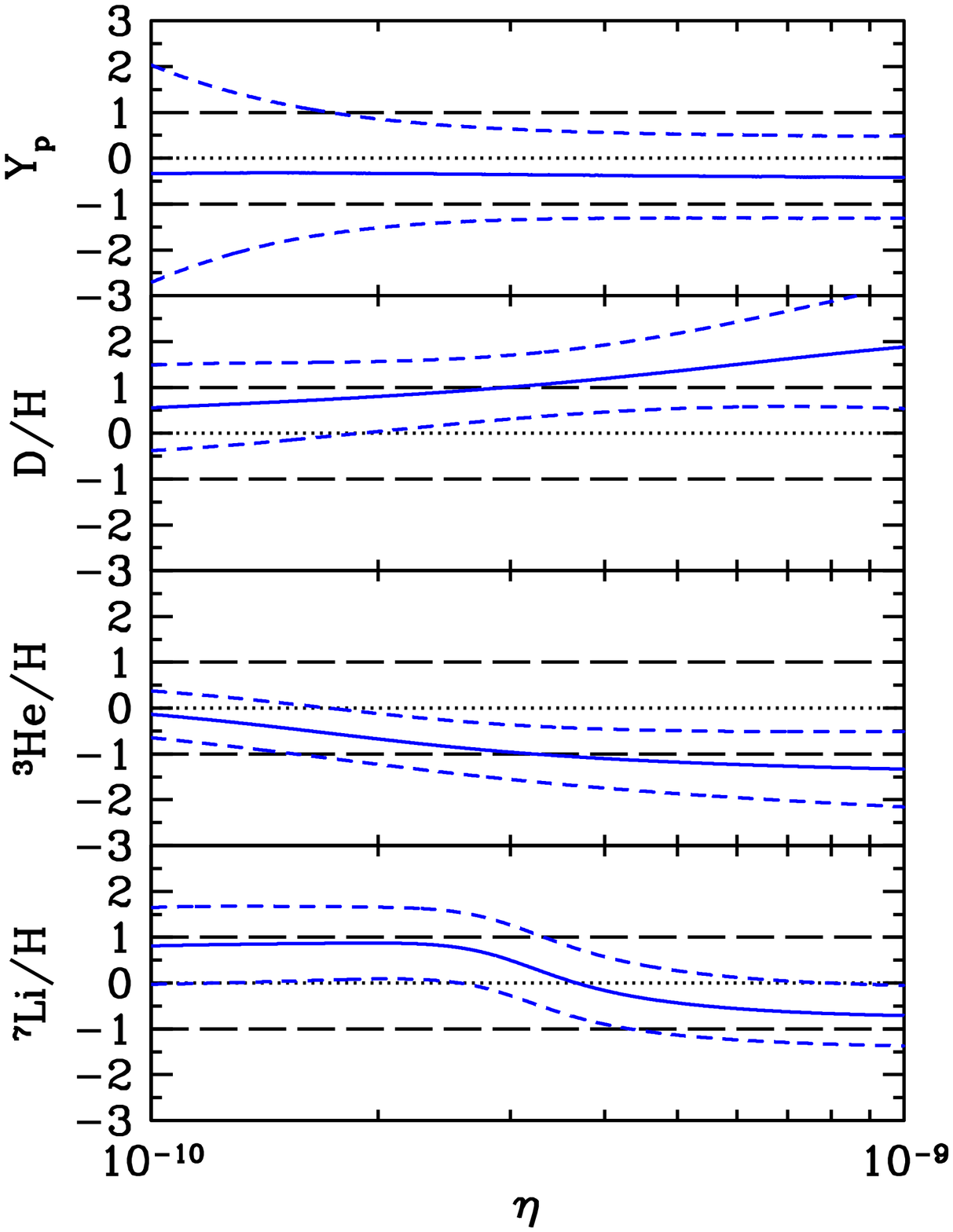,width=3.0in}
\caption{This figure shows the difference between light element yields
using this compilation and that of Cyburt, Fields and Olive
\cite{cfo1}.  The solid and short-dashed curves show the Cyburt,
Fields and Olive yield means and standard deviations with respect to
this compilation's means and standard deviations, seen here as zero
and $\pm 1$, respectively.
\label{fig:bbncompA}}
\end{center}
\end{figure}

Since Cyburt, Fields and Olive and Nollett and Burles both compare
directly to Smith, Kawano and Malaney showing rough agreement, we
choose to compare only to the former two compilations.  Plotted in
figure~\ref{fig:bbncompA} is the residual between Cyburt, Fields and
Olive~\cite{cfo1} and this compilation, where zero and $\pm 1$
represent the means and standard deviations of this compilation.  The
\he4 mass fraction ($Y_p$) is in good agreement with this
compilation. The \he4 abundance error is slightly increased due to the
inclusion of the uncertainty in Newtion's G${}_N$.  This compilations
treatment of the data, leads to differences for the D, \he3 and \li7
yields.  On the high $\eta$ side ($\eta\ga 3\times10^{-10}$) the
changes are due to the $d(p,\gamma)\he3$ reaction.  The new cross
section is larger than the one determined by the NACRE-based
compilation~\cite{nacre} of Cyburt, Fields and Olive, causing a
subsequent drop in D yields with a simultaneous jump in \he3 and \li7
yields.  For low values of $\eta$, only the mean value of
\li7 is significantly different.  This is due entirely to a slightly
lower cross section for the $t(\alpha,\gamma)\li7$ reaction used here.
The errors of the Cyburt, Fields and Olive compilation are generally
smaller than this compilation's errors.  This is due to the fact that
this compilation has an additional intrinsic normalization error added
in quadrature with the discrepancy normalization error.

\begin{figure}[ht]
\begin{center}
\psfig{file=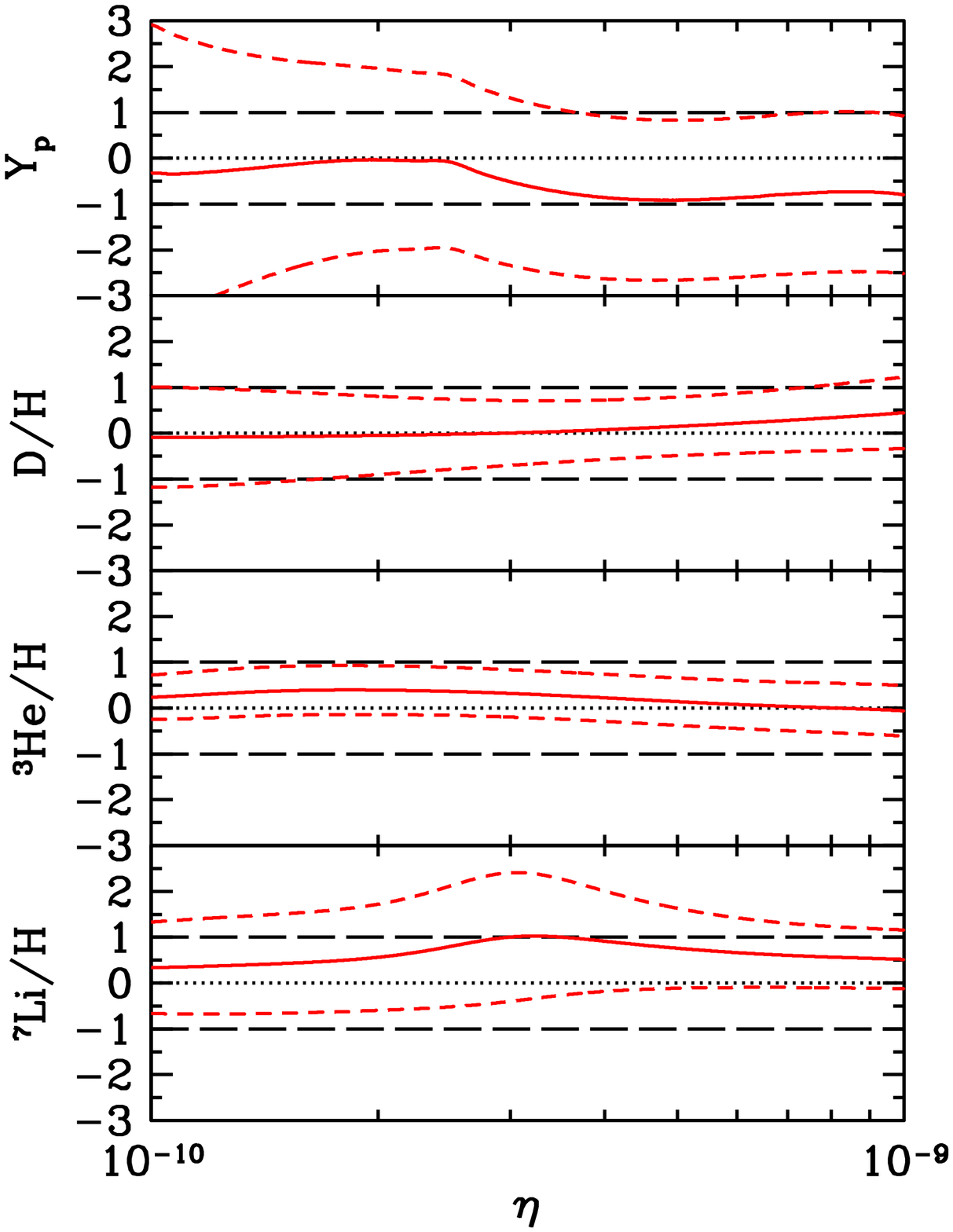,width=3.0in}
\caption{This figure shows the difference between light element yields
using this compilation and that of Nollett and Burles
\cite{nb00}.  The solid and short-dashed curves show the Nollett and
Burles yield means and standard deviations with respect to this
compilation's means and standard deviations, seen here as zero and
$\pm 1$, respectively. 
\label{fig:bbncompB}}
\end{center}
\end{figure}

Similarly, plotted in figure~\ref{fig:bbncompB} is the residual
between Nollett and Burles~\cite{nb00} and this compilation.  The
central values of the abundance yields are in good agreement.  The
most noticeable difference is in the \li7 yields, with this
compilation having slightly lower values.  The high $\eta$ difference
is due to differences in the reactions $d(p,\gamma)\he3$ and
$\he3(d,p)\he4$, while on the low $\eta$ side the differences are
primarily due to the reaction $t(d,n)\he4$.  The errors of Nollett and
Burle's compilation are comparable, but generally smaller than this
compilation's errors, because of the statistical nature of their
errors.  The larger \he4 error is due to the larger uncertainty in the
neutron lifetime adopted by Nollett and Burles.

It is reassuring that this new BBN nuclear compilation agrees quite
well with the previous studies of Smith, Kawano and
Malaney~\cite{skm}, Nollett and Burles~\cite{nb00} and Cyburt, Fields
and Olive~\cite{cfo1}, as well as the two more recent calculations by
Cuoco {\em et al}.~\cite{cimm} and Coc {\em et al}.~\cite{cvdaa},
though the new compilations do not present rate representations that
can be compared directly.  This work's rigorous treatment of
systematic uncertainties also suggests that these previous works may
have underestimated the true error budget in the light element
abundance predictions.  This comparison also suggests where new
nuclear data will be most useful.  To illustrate this point, scalings
are created that explicitly show how the baryon-to-photon ratio
$\eta$, Newton's G and the reaction rates affect the light element
abundance predictions.  These scalings are calculated numerically, by
finding the logarithmic derivatives of the light element abundance
predictions with respect to parameters and key reaction rates,
relative to a fiducial model where
$\eta=6.14\times10^{-10}$~\cite{flsv}. They can be used to either
predict light element abundances or propagate uncertainties, but these
scalings are only approximate and will change for models with $\eta$
very far from its fiducial value.  We use them here, only to discuss
how BBN's predictions depend on the various inputs.  The nuclear
reactions are parametrized here through $R_i$, where $i$ refers to the
subsection number assignment for that reaction in
sect.~\ref{sect:cs_results} (i.e. $R_2$, $R_4$, and $R_5$  correspond
with the $p(n,\gamma)d$, $d(d,n)\he3$, and $d(d,p)t$ reactions
respectively).  The $R_i$ can be thought of as reaction
normalizations, such that the current compilation is $R_i=1.0$.
The scalings are:
\beqar
\label{eqn:scal-he4}
Y_p &=& 0.24849 \left( \frac{10^{10}\eta}{6.14}\right)^{\!\!0.39}\!\!\left(\frac{\tau_n}{\tau_{n,0}}\right)^{\!\!0.72}\!\!\left(\frac{{\rm G}_N}{{\rm G}_{N,0}}\right)^{\!\!0.35}\!\!\!\!\!\! R_4^{0.006} R_5^{0.005} R_2^{0.005} \\
\label{eqn:scal-d}
10^5\frac{\rm D}{\rm H} &=& 2.558\left(\frac{10^{10}\eta}{6.14}\right)^{\!\!-1.62}\!\!\!\left(\frac{\tau_n}{\tau_{n,0}}\right)^{\!\!0.41}\!\!\left( \frac{{\rm G}_N}{{\rm G}_{N,0}}\right)^{\!\!0.95}\!\!\!\!\!\!  R_4^{-0.55} R_5^{-0.45} R_3^{-0.32} R_2^{-0.20} \\
\label{eqn:scal-he3}
10^6\frac{\he3}{\rm H} &=& 10.086\left( \frac{10^{10}\eta}{6.14}\right)^{\!\!-0.59}\!\!\left(\frac{\tau_n}{\tau_{n,0}}\right)^{\!\!0.15}\!\!\left(\frac{{\rm G}_N}{{\rm G}_{N,0}}\right)^{\!\!0.34}\!\!\!\!\!\! R_8^{-0.77} R_3^{0.38} R_5^{-0.25} R_4^{-.20} R_6^{-0.17} R_2^{0.08}  \\
\label{eqn:scal-li7}
10^{10}\frac{\li7}{\rm H} &=& 4.364\left( \frac{10^{10}\eta}{6.14}\right)^{\!\!2.12}\!\!\left(\frac{\tau_n}{\tau_{n,0}}\right)^{\!\!0.44}\!\!\left(\frac{{\rm G}_N}{{\rm G}_{N,0}}\right)^{\!\!-0.72}\!\!\!\!\!\! R_2^{1.34} R_9^{0.96} R_8^{-0.76} R_{11}^{-0.71} R_4^{0.71} R_3^{0.59} R_6^{-0.27}.
\eeqar 

As clearly seen in the scalings, $Y_p$ is dominated by the neutron
mean lifetime and Newton's G${}_N$, while the reactions $d(d,n)\he3$,
$d(d,p)t$ and $p(n,\gamma)d$ only slightly change the predictions.
The dramatic drop in sensitivity (scaling powers $\sim0.5$ to
$\sim0.005$) is seen in the other light element scalings, meaning
those reactions do not contribute significantly to the overall theory
predicions. Thus, for brevity these reactions are left out of the
scalings for D, \he3 and \li7.  For an accurate D prediction,
$d(d,n)\he3$ and $d(d,p)t$ are key, with $d(p,\gamma)\he3$ following
close behind.  \he3 is the least sensitive to which nuclear
compilation is used, though improvements in its prediction propagates
into an improved \li7 prediction.  For high baryon densities
($\eta\ga3\times10^{-10}$), the reactions $\he3(\alpha,\gamma)\be7$
and $\he3(d,p)\he4$ dominate the \li7 predictions, whilst for low
baryon densities ($\eta\la3\times10^{-10}$) their mirror reactions are
dominant, $t(\alpha,\gamma)\li7$ and $t(d,n)\he4$.  With the precision
of $p(n,\gamma)d$ being $\la2.5$\%, it only enters at the percent or
sub-percent level in the light element prediction uncertainties, and
thus is not the dominant error.  With this new nuclear compilation and
its error budget we are well poised to test the overall concordance
between primordial nucleosynthesis' predictions, the observations of
the light element abundances and of the CMB anisotropy.


\section{Discussion}
\label{sect:disc}

\subsection{Light Element Observations}
With the light element predictions of D, \he3, \he4 and \li7 in hand,
we set out to compare them directly to observations.  
Deuterium is measured in high-redshift QSO absorption line systems via its
isotopic shift from hydrogen.  Under the well-founded assumption that the only
significant astrophysical source of deuterium is the big bang
\cite{epstein}, one can estimate that the amount of D depletion in
these high-shift systems to be less than 1\%.  Thus, making D nearly
primordial and a direct probe of big bang nucleosynthesis.  In several
absorbers of moderate column density (Lyman-limit systems), D has been
observed in multiple Lyman transitions.  We adopt the two deuterium
values from Kirkman {\em et al}.~\cite{kirkman}, one being the world
average of the 5 best deuterium measurements including both single and
multiple absorption systems
\cite{burlestytler,omeara,kirkman,pettini} : 
\beq
\label{eq:D_p_wa}
\pfrac{\rm D}{\rm H}_p = (2.78^{+0.44}_{-0.38}) \times 10^{-5}.
\eeq
and second, the average of the 2 multiple absorption line
systems \cite{omeara,kirkman}; 
\beq
\label{eq:D_p_mls}
\pfrac{\rm D}{\rm H}_p = (2.49^{+0.20}_{-0.18}) \times 10^{-5}.
\eeq

As noted in Kirkman {\em et al}., the $\chi^2$ per degree of freedom
is rather poor for the world average D value ($\chi^2_\nu=4.1$).  Many
possibilities exist that can explain this poor $\chi^2$,
underestimated errors, correlations with column
density, and other systematics~\cite{foscv}.  However,
since we are only dealing with 5 systems, any of these conclusions can be
reached.   The 2 multiple absorption line systems
agree quite well with each other, however this could also be due to
low number statistics.  Future observations will help address these
concerns.

Unlike D, \he4 is made in stars, and thus co-produced with heavy elements.
Hence the best sites for determining the primordial \he4 abundance are in
metal-poor regions of hot, ionized gas in nearby external galaxies
(extragalactic H{\small II} regions). Helium indeed shows a linear
correlation with metallicity in these systems, and the extrapolation to
zero metallicity gives the primordial abundance (baryonic mass
fraction, $Y_p = \rho_{\he4}/\rho_{\rm
B}$)~\cite{peim,oss,fieldsolive,izotovthuan}. We cite the 2
following values, as have~\cite{cfo1,cfo2,cfo3}
\beq
Y_p = 0.238 \pm 0.002 \pm 0.005.
\label{he4_fo}
\eeq
\beq
Y_p = 0.244 \pm 0.002 \pm 0.005.
\label{he4_it}
\eeq
determined by Fields and Olive (1998)~\cite{fieldsolive}, and Izotov
and Thuan (1998)~\cite{izotovthuan}, respectively from a large body of
data representing dozens of extragalactic H{\small II} regions.  The
difference between the two values is due primarily to adopted analysis
techniques treating \he4 emission lines and underlying stellar
absorption, as most of the systems are the same between the two sets. Here,
the first error is statistical and reflects the large sample of
systems, whilst the second error is systematic and dominates.  Since
Izotov and Thuan do not quantify a systematic error, we adopt the
systematic error discussed in Fields and Olive and explored in Olive and
Skillman (2001)~\cite{OSk}.  As suggested by Olive and Skillman, these
systematics need to be further explored.

Helium-3 is observed as well, but through its hyperfine emission in
the radio band, limiting observations to Galactic H{\small II}
regions.  The sample size of the \he3 data is rather sparce and
localized around a fairly narrow band in metallicities~\cite{bbr}.
Combined with a considerable dispersion, a model independent
determination of the primordial \he3 abundance is prohibitive.  The
Galactic evolution of \he3 is also poorly understood, as it is not
known if \he3 increases or decreases from its primordial
value~\cite{vofc}, manefesting itself as a large extrapolation error
in model-dependent approaches.  We thus, do not use \he3 observations
to probe primordial nucleosynthesis.

The primordial \li7 abundance is determined from observations of old
metal-poor stars, particularly those in the Galactic stellar halo
(Population II).  For very low metallicities, the \li7 abundance is
found to be nearly constant, the so-called ``Spite
Plateau''~\cite{spite}.  From this, a primordial abundance is
inferred.  An analysis of a set of Pop. II stars with high signal to
noise data was performed by Ryan {\em et al}.~\cite{ryan}, taking into
account various chemical and stellar evolution effects.  Their
primordial \li7 abundance is:
\beq
\label{ryan}
\pfrac{\li7}{\rm H}_p = \EE{(1.23 \pm 0.06_{-0.32}^{+0.68})}{-10}\ ({\rm
95\% \ CL}),
\eeq
where the small statistical error is overshadowed by systematic
uncertainties.  A recent determination by Bonifacio {\em et
al}.~\cite{bonifacio}, based on observations of stars in a globular
cluster, yields slightly different results.  The difference is mainly
attributable to the different methods used to callibrate stellar
atmosphere parameters, in particular the effective temperature.  Their
analysis yields ${\li7/{\rm H}}_p = (2.19^{+0.46}_{-0.38}) \times 10^{-10}$.
The difference between these numbers is a measure of the systematic
error, which has apparently been underestimated by Ryan {\em et al}..
We thus adopt both observations for use as probes of primordial
nucleosynthesis.

Since standard primordial nucleosynthesis is a one parameter theory,
depending on the baryon-to-photon ratio, $\eta$ or equivalently the
baryon density, we can use light element abundance determinations to
measure the baryon content of the universe.  We discuss the
implications of adopting the observations mentioned, on BBN
concordance.

\begin{figure}[ht]
\begin{center}
\psfig{file=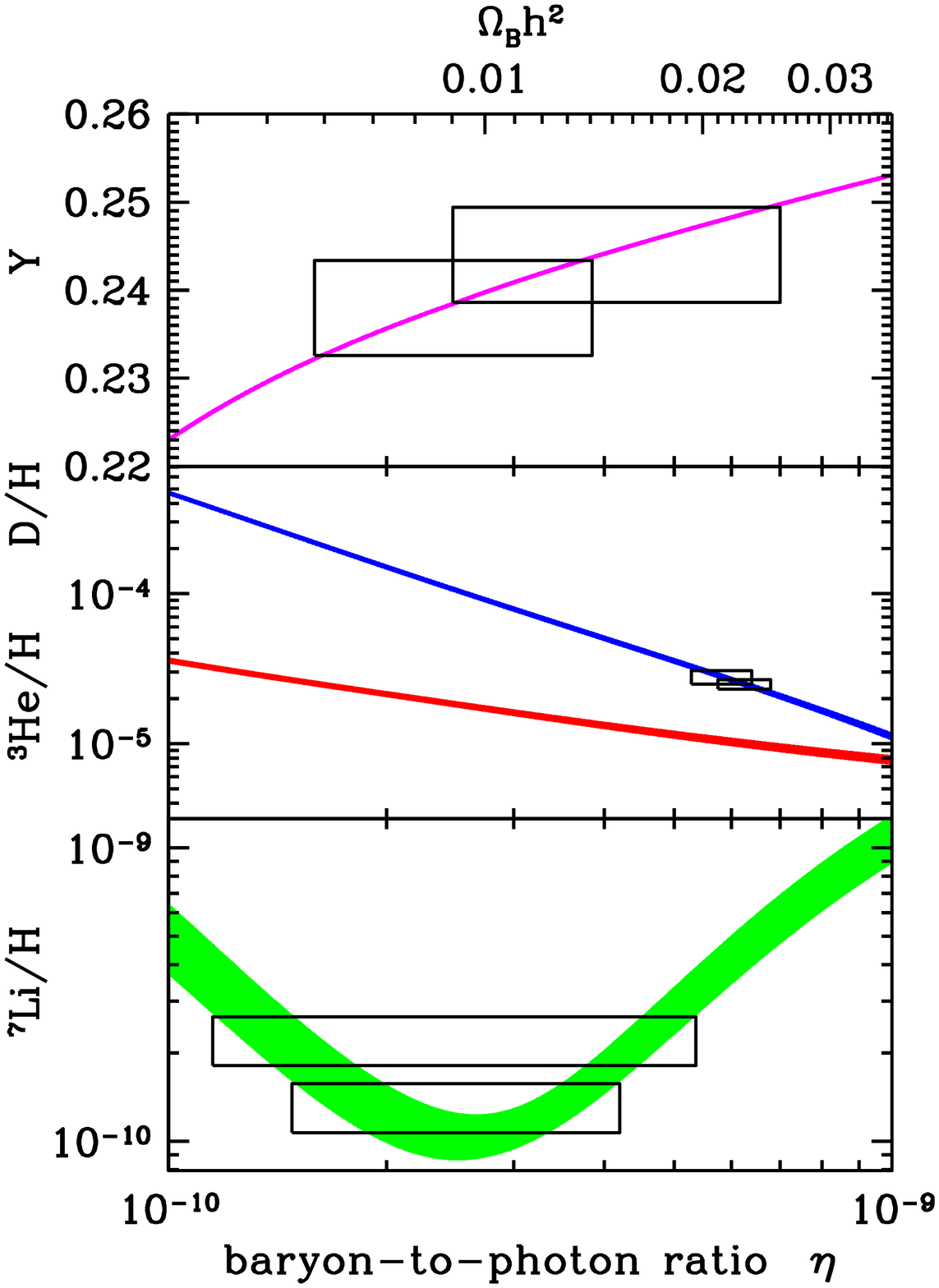,width=2.75in}
\caption{\label{fig:4-abeta_lec}
Shown in the figure are the light element predictions.  The mass
fraction of \he4 ($Y_p$) and the mole fractions relative to hydrogen,
${\rm D/H}$, $\he3/{\rm H}$ and $\li7/{\rm H}$ are plotted against the
baryon-to-photon ratio.  The width of each curve represents the 1
$\sigma$ or 68\% confidence errors in the light element predictions.
The outlined boxes represent the light element observational
constraints on the baryon density.  }
\end{center}
\end{figure}

Shown in fig.~\ref{fig:4-abeta_lec} are the light element predictions
with outlined boxes showing the observational constraints and the
$\eta$ ranges allowed by each.  There is no value of baryon density for
which any three abundance observations agree well, as seen
quantitatively in tab.~\ref{tab:barab}.  Treating all observations
equally, we can only reliably constrain the baryon-to-photon ratio to
lie between $1 \la 10^{10}\eta \la 7$.  There is only marginal
agreement at the 95\% confidence level.  This marginal concordance is
also evaluated in previous works
\cite{cfo1,cfo2} with the use of theory and observationally based
likelihoods \cite{ostys,ot97,ot99,lisi}.  If we limit ourselves to D
only constraints we find that $10^{10}\eta=6.28^{+0.34}_{-0.35}$ and
$5.92^{+0.55}_{-0.58}$, for the multiple absorption and world averages
respectively.

\begin{table}[ht]
\begin{center}\caption{This table lists the baryon density constraints 
placed by various light element observations using this works theory
predictions.  For comparison, the WMAP team's result for the baryon
density is also shown.  The numbers cited are the mostly likely values
and their respective 68\% central confidence limits.  Since the
\cite{bonifacio} \li7 constraint lies above the ``dip'' in the theory
prediction, it has two distinct predictions for the baryon density, a
low baryon density constraint $I$ and a high baryon density constraint
$II$. }
\label{tab:barab}
\begin{tabular}{||c|c|c||}
\hline\hline
Observations & \ \ $\eta_{10}\equiv 10^{10}\eta$ \ \ & $\Omega_{\rm B}h^2$ \\
\hline\hline
D/H = $(2.49^{+0.20}_{-0.18})\times10^{-5}$~\cite{omeara,kirkman} & $6.28^{+0.34}_{-0.35}$ & $0.0229\pm0.0013$ \\
\hline
D/H = $(2.78^{+0.44}_{-0.38})\times10^{-5}$~\cite{burlestytler,omeara,kirkman,pettini} 
 & $5.92^{+0.55}_{-0.58}$ & $0.0216^{+0.0020}_{-0.0021}$ \\
\hline\hline
$Y_p = 0.238\pm 0.002 \pm 0.005$~\cite{fieldsolive} & $2.39^{+1.75}_{-0.87}$ & $0.0087^{+0.0064}_{-0.0031}$ \\
\hline
$Y_p = 0.244\pm 0.002 \pm 0.005$~\cite{izotovthuan} & $3.95^{+3.54}_{-1.64}$ & $0.0144^{+0.0129}_{-0.0060}$ \\
\hline\hline
\li7/H = $(1.23\pm 0.03 ^{+0.34}_{-0.16})\times10^{-10}$\cite{ryan} & $3.19^{+0.41}_{-1.23}$ & $0.0116^{+0.0015}_{-0.0044}$ \\
\hline
\li7/H = $(2.19^{+0.46}_{-0.38})\times10^{-10}$\cite{bonifacio} & $^{I}1.49^{+0.25}_{-0.22}$ & $0.0055^{+0.0009}_{-0.0008}$ \\
 ``  \ \ \    `` & $^{II}4.41^{+0.57}_{-0.51}$ & $0.0161^{+0.0021}_{-0.0019}$ \\
\hline\hline
WMAP (2003)~\cite{wmap} & $6.14\pm0.25$ & $0.0224\pm0.0009$ \\
\hline\hline
\end{tabular}
\end{center}
\end{table}

This tension could either be pointing out unknown systematics in the
abundance observations, or be telling us that there is new physics to
be learnt.  We address both of these in the following sections.  An
independent measure of the baryon density will eliminate it as a free
parameter for BBN.  This independent determination will act as a
tie-breaker among light element observations and lead the way to
understanding this tension, whether the disagreement results from
underlying systematics or new physics.

\subsection{Observational Concordance}
\begin{figure}[ht]
\begin{center}
\psfig{file=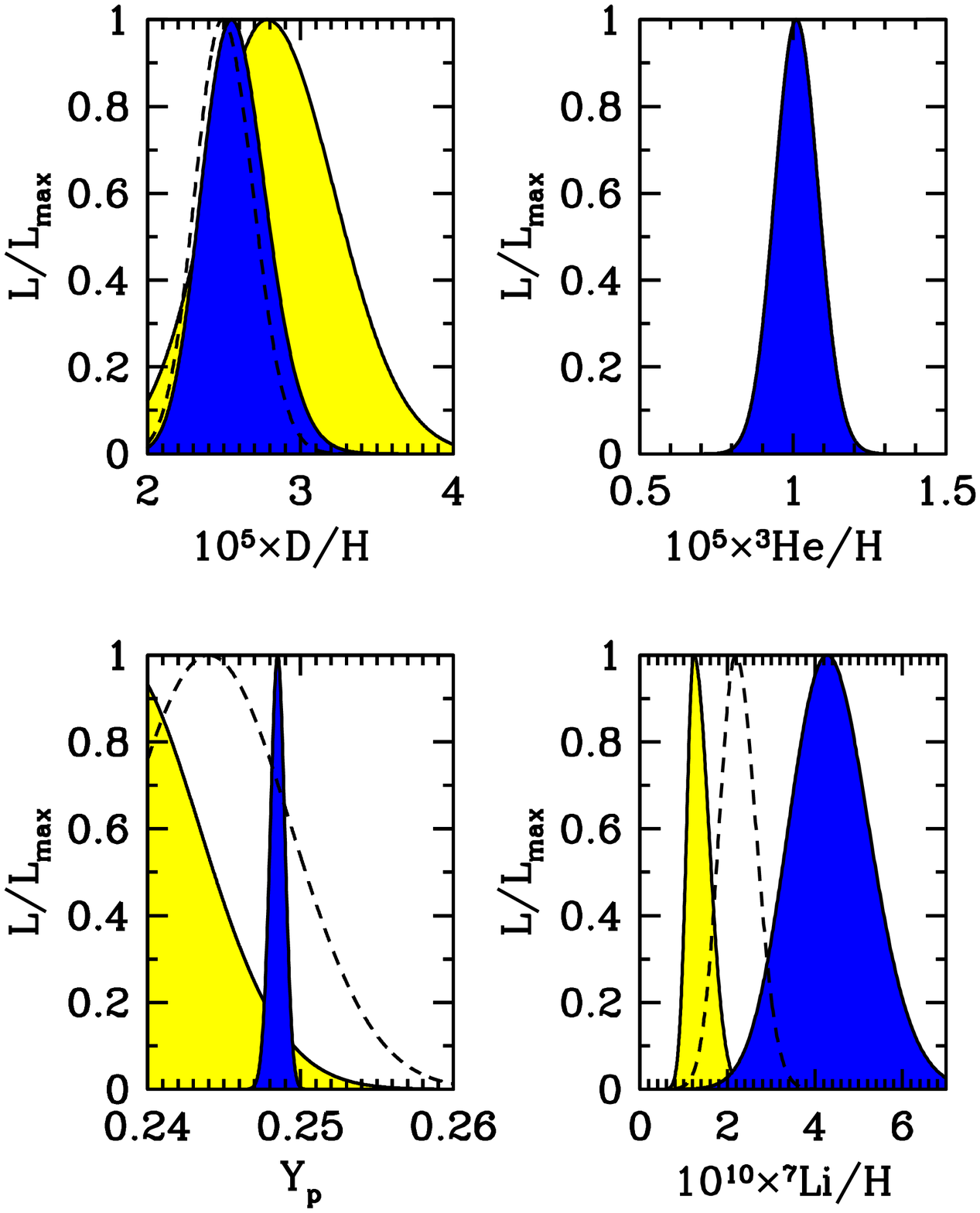,width=2.75in}
\caption{Primordial light element abundances as predicted by BBN and
WMAP (dark shaded regions).  Different observational assessments of
primordial abundances are plotted as follows: (a) the light shaded
region shows ${\rm D/H} = (2.78^{+0.44}_{-0.38})\times
10^{-5}$\cite{burlestytler,omeara,kirkman,pettini}, while the
dashed curve shows ${\rm D/H} = (2.49^{+0.20}_{-0.18})\times
10^{-5}$\cite{omeara,kirkman}; (b) no
observations plotted; see text (c)  the light shaded region shows
$Y_p=0.238\pm0.002\pm0.005$ \cite{fieldsolive}, while the dashed curve
shows $Y_p = 0.244\pm0.002\pm0.005$ \cite{izotovthuan}; (d) the light
shaded region shows $\li7/{\rm H} = (1.23^{+0.34}_{-0.16})\times 10^{-10}$
\cite{ryan}, while the dashed curve shows $\li7/{\rm H} =
(2.19^{+0.46}_{-0.38})\times 10^{-10}$ \cite{bonifacio}. 
\label{fig:4-LYi_WMAP}}
\end{center}
\end{figure}

As mentioned at the beginning of the work, the CMB anisotropies detail
information about the shape and content of our universe.  With the
first data release of the WMAP team \cite{wmap}, several cosmological
parameters have been measured to unprecedented accuracy, including the baryon
density, which is measured to be, 
\beq
\label{eqn:4.1}
\Omega_{\rm B}h^2 = 0.0224\pm 0.0009.
\eeq
This corresponds with a baryon-to-photon ratio of $\eta = (6.14 \pm
0.25)\times 10^{-10}$.  This is a 4\% measurement, which makes it a
sharper baryon probe than any light element currently is.  Since we no
longer are required to use the light element abundances to tell us the
baryon content of the universe, the analysis completely changes.  Now
we can predict the light element abundances, with this baryon density
and compare those predictions with the light element observations.
With WMAP's baryon density we get:
\beqar
Y_p &=& 0.2485\pm 0.0005 \\
{\rm D/H} &=& \left( 2.55^{+0.21}_{-0.20}\right)\times 10^{-5} \\
\he3/{\rm H} &=& \left( 10.12^{+0.67}_{-0.66}\right)\times 10^{-6} \\
\li7/{\rm H} &=& \left( 4.26^{+0.91}_{-0.86}\right)\times 10^{-10} 
\eeqar
for the light element predictions with our new nuclear reaction
network.  Figure~\ref{fig:4-LYi_WMAP} shows the predictions and compares
them directly with observations.

In order to quantify the level of concordance, we define an effective $\chi^2$.
\beq
\chi^2_{eff} = \frac{(A_{obs} - A_{wmap})^2}{\sigma_{obs}^2 + \sigma_{wmap}^2},
\eeq
where $A_{obs}$ and $A_{wmap}$ are the most likely values of the light
element abundances for the adopted observations and that predicted
with the WMAP baryon density.  $\sigma_{obs}$ and $\sigma_{wmap}$ are
the corresponding 68\% confidence errors.  The $\chi^2$ values are
shown in tab.~\ref{tab:abchi2}.

\begin{table}[ht]
\begin{center}\caption{This table lists the effective $\chi^2$'s for 
each observational constraint of the light element abundances, given
the WMAP baryon density and this compilation's BBN theory. A
$\chi^2_{eff}$ value smaller than unity means concordance, while a
value large than unity shows discordance.  The magnitude of
discordance is measured by $\sqrt{\chi^2_{eff}}$, a measure of how
many ``$\sigma$'' of discordance exists.}
\label{tab:abchi2}
\begin{tabular}{||c|c|c||}
\hline\hline
Observations & \ \ $\chi^2_{eff}$ \ \ & $\sqrt{\chi^2_{eff}}$ \\
\hline\hline
D/H = $(2.49^{+0.20}_{-0.18})\times10^{-5}$~\cite{omeara,kirkman} & 0.045 & 0.212 \\
\hline
D/H = $(2.78^{+0.44}_{-0.38})\times10^{-5}$~\cite{burlestytler,omeara,kirkman,pettini} 
 & 0.281 & 0.530 \\
\hline\hline
$Y_p = 0.238\pm 0.002 \pm 0.005$~\cite{fieldsolive} & 3.77 & 1.94 \\
\hline
$Y_p = 0.244\pm 0.002 \pm 0.005$~\cite{izotovthuan} & 0.692 & 0.832 \\
\hline
$Y_p = 0.238\pm 0.002$~\cite{fieldsolive} & 25.9 & 5.09 \\
\hline
$Y_p = 0.244\pm 0.002$~\cite{izotovthuan} & 4.77 & 2.18 \\
\hline\hline
\li7/H = $(1.23\pm 0.03 ^{+0.34}_{-0.16})\times10^{-10}$\cite{ryan} & 10.72 & 3.27 \\
\hline
\li7/H = $(2.19^{+0.46}_{-0.38})\times10^{-10}$\cite{bonifacio} & 4.50 & 2.12 \\
\hline\hline
\end{tabular}
\end{center}
\end{table}

As one can see, the two adopted observational values of deuterium
agree with the BBN+WMAP prediction, both having $\chi^2_{eff}$'s
smaller than unity.  It is unclear if the slightly worse $\chi^2$ of
the world average is due to unknown systematics or just poor
statistics.  Hopefully, with future automated searches, many more of
these special absorption systems can be found.  It is interesting to
note that the WMAP baryon density contributes significantly to the
uncertainty in the predicted D abundance.  Future CMB experiments will
reduce this uncertainty, at which time the BBN nuclear uncertainties
will totally dominate the theory predictions.  Thus motivating renewed
efforts for new cross section measurements.

The Izotov and Thuan value, $Y_p=0.244$~\cite{izotovthuan} value
agrees with theory predictions only if the systematic errors are taken
into account as discussed earlier.  If they are ignored, this number
shows discordance at more than the 2-$\sigma$ level.  The Fields and
Olive value, $Y_p=0.238$~\cite{fieldsolive} shows discordance at the
2-$\sigma$ level with systematic uncertainties.  If they are ignored
here, the discordance becomes a 5-$\sigma$ deviation.  It is clear
that a more detailed study of these systematics, including the effects
of underlying stellar absorption and varying treatments of emission
lines, is needed~\cite{OSk}.  One may also consider the new
evaluation by Izotov and Thuan~\cite{newIT}, finding $Y_p=0.2421\pm
0.0021$, a nearly 3-$\sigma$ deviation with this compilation's CMB+BBN
predictions.  We believe a proper accounting of the systematic errors
will alleviate this discordance.

The CMB itself, is also sensitive to the value of $Y_p$.  First
attempts at constraining $Y_p$ have been performed by Trotta and
Hansen~\cite{trothan} and by Huey, Cyburt and
Wandelt~\cite{hcw}. Future parameter studies, should include $Y_p$ as
a free parameter, rather than adopting the canonical value of 0.24.
The CMB constraint offers an independent determination, free of the
systematics plaquing the determination from extra-galactic H{\small
II} regions.  It also is a direct probe of $Y_p$, so no extrapolations
to zero metallicity are needed for the determination, just high
precision CMB anisotropy data.  With this data, we can use the
CMB-determined $Y_p$ to quantify the level of observational
systematics discussed above.  By combining BBN predictions with CMB
observations, we can also learn about stellar
evolution~\cite{bono,cassisi}.  One should also be mindful of the
baryon density dependence on $Y_p$ given in eqn.~\ref{eqn:baryeta},
especially when combining BBN and CMB results.

The WMAP+BBN prediction for \li7 disagrees with both
observationally-based primordial \li7 abundances, with the Ryan {\em
et al}.~\cite{ryan} and Bonifacio {\em et al}.~\cite{bonifacio}
numbers showing discordance at the 3 and 2-$\sigma$ level.  As already
mentioned, the difference between these two sets of observations is a
measure of the systematic error due to the different methods used.
This is not large enough to account for all of the discrepency between the
observation-based and predicted values. An often discussed possibility
is the depletion of atmospheric \li7.  This possibility faces the
strong constraint that the observed lithium abundances show extremely
little dispersion, making it unlikely that stellar processes which
depend on the temperature, mass, and rotation velocity of the star all
destroy \li7 by the same amount.  Uniform depletion
factors of order 0.2 dex (a factor of 1.6) have been discussed
\cite{dep}. It is clear that either (or both) the base-line abundances
of \li7 have been poorly derived or stellar depletion is far more
important than previously thought.  Of course, it is possible that if
systematic errors can be ruled out, a persistent discrepancy in \li7
could point to new physics.

\subsection{Implications for non-standard BBN}

With the goal of maintaining concordance with observations, we examine
how sharply we can deviate from the standard model.  Often the effect of
new physics can be parameterized in terms of additional relativistic
degrees of freedom during the epoch of primordial nucleosynthesis,
usually expressed in terms of the effective number of neutrino
species, $N_{\nu,eff}$.  Traditionally, D or \li7 observations were
used to fix the baryon density and the \he4 mass fraction was used to
fix $N_{\nu,eff}$.  These limits are thoroughly described elsewhere
\cite{cfo2,ssg,flsv,dolgov}.  Moreover, as we have noted, the
observed \he4 appears lower than the WMAP+BBN value.  This discrepancy
is likely due to systematic errors, but could point to new physics.
Until this situation is better understood, caution is in order.
Fortunately, in the post-WMAP era, we can now use the CMB-determined
baryon density (eqn.~\ref{eqn:4.1}), to remove it as a free parameter
from BBN theory and use any or all abundance observations to constrain
$N_{\nu,eff}$ \cite{cfo2,cfo3,cimm,hannestad,barger}.  In particular, we have
computed the likelihood distributions for $N_{\nu,eff}$ using the WMAP
$\eta$ and several of the light element observations.

\begin{figure}[ht]
\begin{center}
\psfig{file=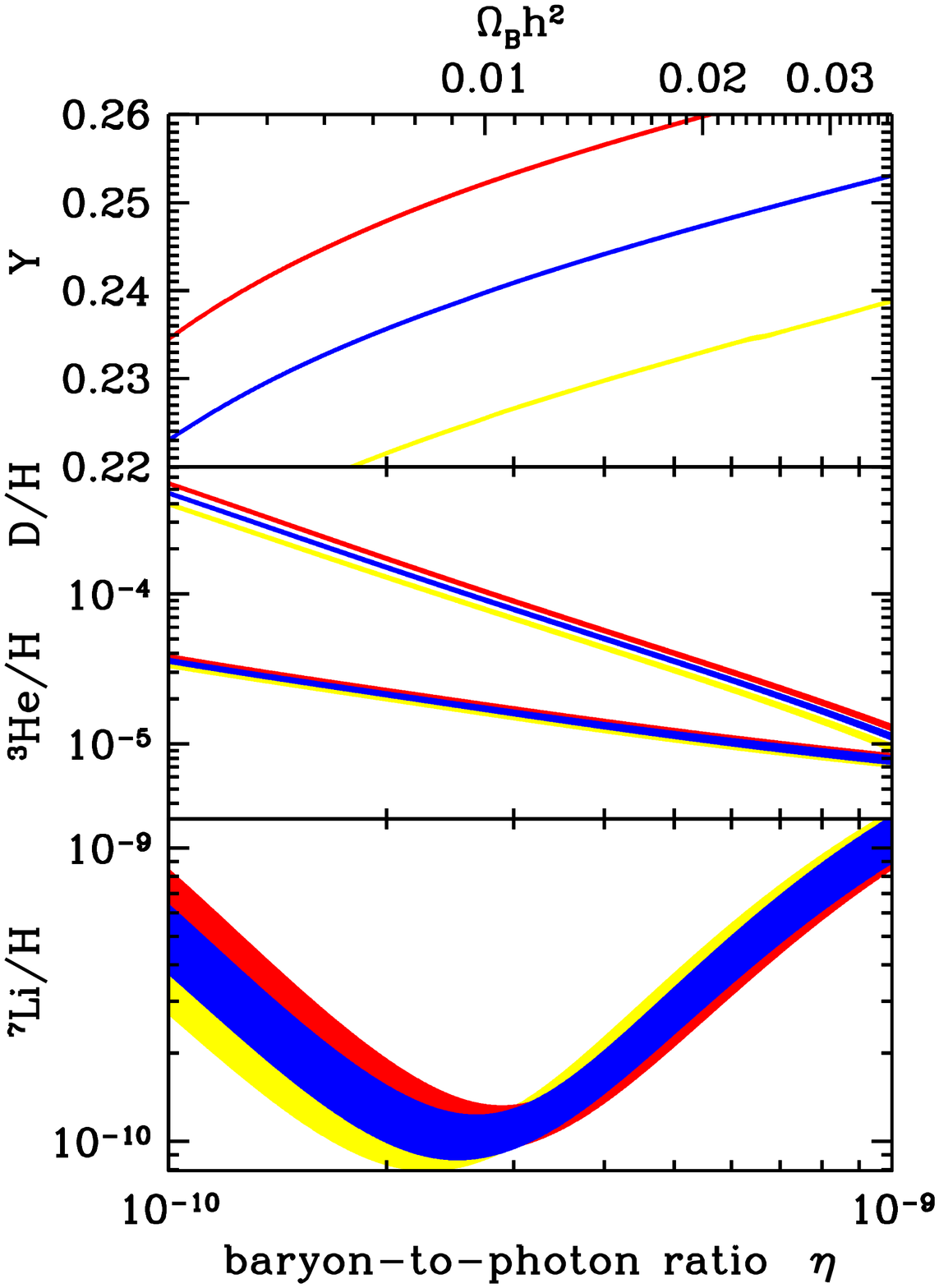,width=2.75in}
\caption{The light element predictions plotted against the
baryon-to-photon ratio for different values for $N_{\nu,eff}$.  The
light shaded region corresponds with $N_{\nu,eff}=2.0$, the medium
shaded with $N_{\nu,eff}=4.0$ and the dark shaded with $N_{\nu,eff}=3.0$.
\label{fig:4-abeta_nu}}
\end{center}
\end{figure}

To first gauge what elements are sensitive to $N_{\nu,eff}$, we have
plotted the primordial abundance predictions for the standard case,
$N_{\nu,eff}=3$, and two non-standard cases, $N_{\nu,eff}=2, 4$ in
fig.~\ref{fig:4-abeta_nu}.  As readily apparent, \he4 is the most
sensitive element.  If we understood the underlying systematics with
the \he4 observations better, this would be the ideal choice for
picking an observation to make the constraint.  However, since we are
unsure about \he4, we must move to another observation.  We see that D
is the next most sensitive, and most notably, with this new
compilation, the differences between the 3  $N_{\nu,eff}$'s are
clearly resolvable.  With current theory and observation uncertainties,
\li7 is not very sensitive to the relativistic
degrees of freedom at the high baryon densities that the CMB prefers,
thus making it not suitable for this analysis.

To gauge the kinds of constraints we can place, we calculate both
\he4 and D constraints seen in tab.~\ref{tab:nnuab}.  The \he4 
observations that appear systematically low in the standard case,
$N_{\nu,eff}=3$, pull down the most likely value of $N_{\nu,eff}$ to
lie between 2 and 3 depending on which abundance we favor.  With our
concerns of systematics in \he4 observations, we do not put much
weight in the \he4-based constraints, but we do note that the Izotov
and Thuan~\cite{izotovthuan} $Y_p$ determination is in fair agreement
with the Standard Model of Particle phyics value of $N_{\nu,eff}=3.0$,
as long as systematic errors are taken into account. However, the
$N_{\nu,eff}=3.0$ D prediction is in accord with the CMB baryon
density, thus yielding less pull away from $N_{\nu,eff}=3.0$ in the
non-standard model.  It is very interesting that each
observation's most likely $N_{\nu,eff}$ lie on opposite sides of the
standard BBN value of 3.  Using the multiple absorption line system
average~\cite{omeara,kirkman}, we find
$N_{\nu,eff}=2.78^{+0.87}_{-0.76}$.  With the world average, we find
$N_{\nu,eff}=3.65^{+1.46}_{-1.30}$.  Even though D's dependence on
$N_{\nu,eff}$ is smaller than its dependence on the baryon density, a
sufficiently accurate measurement (e.g. WMAP) of $\eta$, will help
make D a more accurate probe.  Thus demanding both improved nuclear
data and more D observations.

\begin{table}[ht]
\begin{center}\caption{This table lists the $N_{\nu,eff}$ constraints 
placed by various light element observations using this works theory
predictions and the WMAP team's baryon density, $\Omega_{\rm B}h^2 =
0.0224\pm0.0009$~\cite{wmap}.  The numbers cited are the mostly likely
values and their respective 68\% central confidence limits. }
\label{tab:nnuab}
\begin{tabular}{||c|c||}
\hline\hline
Observations & $N_{\nu,eff}$ \\
\hline\hline
D/H = $(2.49^{+0.20}_{-0.18})\times10^{-5}$~\cite{omeara,kirkman} & $2.78^{+0.87}_{-0.76}$  \\
\hline
D/H = $(2.78^{+0.44}_{-0.38})\times10^{-5}$~\cite{burlestytler,omeara,kirkman,pettini} 
 & \ \ $3.65^{+1.46}_{-1.30}$ \ \ \\
\hline\hline
$Y_p = 0.238\pm 0.002 \pm 0.005$~\cite{fieldsolive} & \ \ $2.26^{+0.37}_{-0.36}$ \ \  \\
\hline
$Y_p = 0.244\pm 0.002 \pm 0.005$~\cite{izotovthuan} & \ \ $2.67^{+0.40}_{-0.38}$ \ \  \\
\hline\hline
\end{tabular}
\end{center}
\end{table}


\section{Conclusions}
\label{sect:concl}

Primordial nucleosynthesis has entered a new era.
With the precision observations of WMAP,
the CMB has become the premier cosmic baryometer.
The independent BBN and CMB predictions
for $\eta$ are in good agreement (particularly when D is used in BBN),
indicating that cosmology has passed a fundamental test.
Moreover, this agreement allows us to use BBN in a new way, as
the CMB removes $\eta$ as a free parameter.
One can then adopt the standard BBN predictions,
and use $\eta_{\rm CMB}$ to infer primordial abundances;
by comparing these to light element abundances in
different settings, one gains new insight into
the astrophysics of stars, H{\small II} regions, cosmic rays,
and chemical evolution, to name a few examples.
Alternately, WMAP transforms BBN into a sharper
probe of new physics in the early universe;
with $\eta_{\rm CMB}$ fixed, {\em all} of the light elements
constrain non-standard nucleosynthesis, with
$\nnu$ being one example.

As BBN assumes a new role, much work remains to be done.
To leverage the power of the WMAP precision
requires the highest possible precision in light element
observations.
Further improvements in the primordial D abundance
can open the door to D as a powerful probe of early universe physics.
Improved \he3 observations can offer new insight into
stellar and chemical evolution in the Galaxy.
And perhaps most pressing, the WMAP prediction for primordial
\he4 and particularly \li7 are higher than
the current observed abundances; 
it remains to be resolved what systematic effects
(or new physics!) has led to this discrepancy.

This work has always been motivated by the idea of precision
cosmology.  We have laid out a rigorous procedure for determining best
fit parameters and their uncertainties.  We have explicitly taken into
account the correlations among data points and their normalization
errors.  We found it necessary to define two systematic uncertainties,
one is a calculation of the inherent normalization of the data.  The
second is a measure of how well different data sets agree with each
other.  This work generally agrees with previous studies, except in
some special cases as discussed.

Using these updated nuclear inputs, we compute the new BBN
abundance predictions, and quantitatively examine their concordance
with observations.   BBN theory uncertainties are
dominated by the following reactions: $d(d,n)\he3$, $d(d,p)t$,
$d(p,\gamma)\he3$, $\he3(\alpha,\gamma)\be7$ and  $\he3(d,p)\he4$.
Reducing BBN's uncertainties will allow stronger statements
about concordance.  Depending on what deuterium observations are
adopted, one gets the following constraints on the baryon density:
$\Omega_{\rm B}h^2=0.0229\pm0.0013$ or $\Omega_{\rm B}h^2 =
0.0216^{+0.0020}_{-0.0021}$ at 68\% confidence.  If we instead adopt
the WMAP baryon density, we find the following constraints on the
effective number of neutrinios during BBN:
$N_{\nu,eff}=2.78^{+0.87}_{-0.76}$ or
$N_{\nu,eff}=3.65^{+1.46}_{-1.30}$ at 68\% confidence.  Concerns over
systematics in helium and lithium observations limit the confidence of
the constraints derived from this data.  Further exploration of these
systematics, given new observational techniques and more detailed
models, will be most beneficial in understanding and ultimately
reducing their effects.  Deuterium suffers from a small sample size; a
larger sample size will not only improve statistics but also allow the
examination of possible systematics.  With new nuclear cross section data,
light element abundance observations and the ever increasing
resolution of the CMB anisotropy, tighter constraints can be placed on
nuclear and particle astrophysics.

In closing, it is impressive that our now-exquisite understanding of
the universe at $z \sim 1000$ also confirms our understanding of the
universe at $z \sim 10^{10}$.  This agreement lends great confidence
in the soundness of the hot big bang cosmology, and impels our search
deeper into the early universe.

\begin{acknowledgments}
We would like to thank Brian D. Fields for his constant encouragement
and always useful advise.  We would also like to thank my
collaborators John Ellis, Keith A. Olive, Greg Huey and Benjamin
D. Wandelt for their patience, useful discussions and questions;
Yuri Izotov, Byron Jennings, Ken Nollett, Sean Ryan and Jon J. Thaler
for their helpful dialogue; and Gerry Hale for generously providing
his np-capture data.  We would also like to thank the two anonymous
referees, whose comments were most beneficial during the final
preparation of this paper; and Kazuhide Ichikawa, Hugon Karwowski,
Douglas S. Leonard, Robert Scherrer and Pasquale D. Serpico for their
useful comments and questions, with special thanks to Douglas
S. Leonard who graciously provided a copy of the Ganeev
work~\cite{ganeev}.  The work of R.H.C. was supported by the National
Science Foundation Grant AST-0092939, the University of Illinois
Urbana-Champaign Departments of Astronomy and Physics and the Natural
Sciences and Engineering Research Council of Canada.
\end{acknowledgments}

\appendix
\section{Cross Section Fits}

\subsection{$d(p,\gamma)\he3$}
\noindent
$S(E) = 0.2268(1+22.05E+30.77E^2-9.919E^3) {\rm \ eV\ b}$,

\subsection{$d(d,n)\he3$}
\noindent
$S(E) = 0.05067(1+7.534E-4.225E^2+1.508E^3-0.2041E^4) {\rm \ MeV\ b}$

\subsection{$d(d,p)t$}
\noindent
$S(E) = 0.05115(1+4.685E-1.021E^2) {\rm \ MeV\ b}$

\subsection{$\he3(n,p)t$}
\noindent
$R(E) = 6.846\times 10^{8}(1. -0.464743311577E-01E^{1/2} -0.206566636058E+02E \\
+145.303829979E^{3/2} -517.845305322E^2
+1061.59032882E^{5/2} -1232.39931680E^3\\ 
+748.452414743E^{7/2} -184.417975062E^4) {\rm \ cm^3 \
g^{-1} s^{-1}}$

\subsection{$t(d,n)\he4$}
\noindent
$S(E) = 24.19
(1+3.453E-40.16E^2+285.6E^3-596.4E^4+407.1E^5)/(1 + \left(
\frac{E-E_R}{\Gamma_R/2} \right)^2) {\rm \ MeV\ b},$ \\
where $E_R=0.0482$ MeV and $\Gamma_R=0.0806$ MeV.

\subsection{$\he3(d,p)\he4$}
\noindent
$S(E) = 18.52
(1-4.697E+39.53E^2-109.6E^3+130.7E^4-54.83E^5)/(1 + \left( \frac{E-E_R}{\Gamma_R/2} \right)^2) {\rm \ MeV\ b},$\\ 
where $E_R=0.183$ MeV and $\Gamma_R=0.256$ MeV.

\subsection{$\he3(\alpha,\gamma)\be7$}
\noindent
$S(E) = 0.3861(1+0.8195E-2.194E^2+1.419E^3-0.2780E^4) {\rm \ keV\ b}$

\subsection{$t(\alpha,\gamma)\li7$}
\noindent
$S(E) = 0.08656(1+0.6442E-7.597E^2+12.16E^3-5.336E^4) {\rm \ keV\ b}$

\subsection{$\be7(n,p)\li7$}
\noindent
$R(E)=4.7893\times 10^{9}(1 -4.12682044152E^{1/2}+3.10200988738E\\
+15.8164551655E^{3/2} -45.5822669937E^2+54.7133921087E^{5/2}
-34.7483784037E^3\\ +11.3599443403E^{7/2}
-1.49669812741E^4)+1.0553\times 10^{9}/(1.+((E-E_{R,1})/(0.5\Gamma_{R,1}))^2)\\
+2.0364\times 10^{9}/(1.+((E-E_{R,2})/(0.5\Gamma_{R,2}))^2){\rm \ cm^3 \ g^{-1}
s^{-1}}$\\ where $E_{R,1}=0.32$ MeV, $\Gamma_{R,1}=0.20$ MeV,
$E_{R,2}=2.7$ MeV, and $\Gamma_{R,2}=1.9$ MeV.

\subsection{$\li7(p,\alpha)\he4$}
\noindent
$S(E) = 0.06068(1+3.174E-7.586E^2+8.539E^3-3.216E^4) {\rm \ MeV\ b}$

\section{Thermal Rates}
In adopting these rates, it is a good idea to assume the rate constant
at temperatures above which these fits are no longer valid (see
table~\ref{tab:Trnge}).  This prevents the artificial divergence of
the abundances.

\subsection{$p(n,\gamma)d$}
\noindent
$N_A\langle \sigma v\rangle$ = 4.40654e4*(1.+.0457518*t912-2.47101*t9+4.17185*t932\\ 
     -3.44553*t9*t9+1.72766*t9*t932-.546196*t9**3\\ 
     +.106066*t912*t9**3-.0115306*t9**4+.536436e-3*t912*t9**4)

\subsection{$d(p,\gamma)\he3$}
\noindent
$N_A\langle \sigma v\rangle$ = 7.30909e+3*t9m23*ex(-3.7209/t913)\\ 
      *(1.-10.3497*t913+63.4315*t923-209.780*t9\\ 
      +432.557*t943-571.937*t953+497.303*t9*t9\\ 
      -284.936*t943*t9+106.863*t953*t9-25.7496*t9**3\\ 
      +3.81387*t913*t9**3-.313823*t923*t9**3+.0108908*t9**4)

\subsection{$d(d,n)\he3$}
\noindent
$N_A\langle \sigma v\rangle$ = 1.00749e+9*t9m23*ex(-4.2586/t913)
     \\    *(1. -9.59015*t913+65.2448*t923-247.756*t9
     \\    +596.231*t943-941.064*t953+980.076*t9*t9
     \\    -643.032*t9*t943+211.982*t9*t953+29.0491*t9**3
     \\    -66.1847*t913*t9**3+31.6452*t923*t9**3-7.15147*t9**4
     \\    +.372749*t913*t9**4+.208645*t923*t9**4-.0545129*t9**5
     \\    +.00536216*t913*t9**5-.000157984*t923*t9**5
     \\    -.457514e-5*t9**6+2.123592e-9*t913*t9**6)

\subsection{$d(d,p)t$}
\noindent
$N_A\langle \sigma v\rangle$ = 3.91889e+8*t9m23*ex(-4.2586/t913)
     \\    *(1.+.309233*t913-.337260*t923+2.51922*t9
     \\    -2.79097*t943+2.16082*t953-.976181*t9*t9
     \\    +.210883*t943*t9-.0169027*t953*t9+7.845538e-6*t9**3)

\subsection{$\he3(n,p)t$}
\noindent
$N_A\langle \sigma v\rangle$ = 6.84713e+8*(1.-.0171094*t912-2.66179*t9+8.27463*t932
     \\    -14.3898*t9*t9+15.6385*t932*t9-10.3337*t9**3
     \\    +3.80177*t912*t9**3-.599790*t9**4-.0139213*t912*t9**4
     \\    +.0140311*t9**5-.00106709*t912*t9**5+1.06709e-6*t9**6)

\subsection{$t(d,n)\he4$}
\noindent
$N_A\langle \sigma v\rangle$ = 1.78988e12*t9m23*ex(-4.5245/t913)
     \\    /(1. + ((0.129964*t923-0.0482)/(0.5*0.0806))**2)
     \\    *(1.-14.3137899*t913+92.4325675*t923
     \\    -314.645738*t9+641.100355*t943-844.106855*t953
     \\    +752.418564*t9*t9-465.820564*t943*t9
     \\    +202.276143*t953*t9-61.3172473*t9**3
     \\    +12.6913874*t913*t9**3-1.707344*t923*t9**3
     \\    +.134399048*t9**4-.00469341945*t913*t9**4)

\subsection{$\he3(d,p)\he4$}
\noindent
$N_A\langle \sigma v\rangle$ = 5.67897e12*t9m23*ex(-7.1840/t913)
     \\    /(1. + ((0.206357*t923-0.183)/(0.5*0.256))**2)
     \\    *(1.-8.59410908*t913+31.1979775*t923
     \\    -61.2218616*t9+72.0331037*t943-52.8696341*t953
     \\    +23.7371543*t9**2-5.4569107*t943*t9-.226478266*t953*t9
     \\    +.583380161*t9**3-.190978484*t913*t9**3
     \\    +.031949394*t923*t9**3-.00284146599*t9**4
     \\    +.106749198e-3*t913*t9**4)

\subsection{$\he3(\alpha,\gamma)\be7$}
\noindent
$N_A\langle \sigma v\rangle$ = 3.94207e+6*t9m23*ex(-12.8274/t913)
     \\    *(1. + .185267*t913 - .837432*t923
     \\    +7.23019*t9 - 26.1976*t943 +41.6914*t953
     \\    +19.4465*t9*t9-215.248*t9*t943+422.548*t9*t953
     \\    -412.866*t9**3+176.691*t913*t9**3+45.8891*t923*t9**3
     \\    -100.644*t9**4+54.8984*t913*t9**4-14.1903*t923*t9**4
     \\    +1.48464*t9**5)

\subsection{$t(\alpha,\gamma)\li7$}
\noindent
$N_A\langle \sigma v\rangle$ = 4.65494351e+6*t9m23*ex(-8.0808/t913)
     \\    *(1. - 12.3956341*t913 + 76.2717899*t923
     \\    -250.678479*t9+446.413119*t943 -289.008201*t953
     \\    -474.786707*t9*t9 +1346.42142*t9*t943 -1503.09444*t9*t953
     \\     +923.138882*t9**3-306.14089*t913*t9**3+42.9886919*t923*t9**3)

\subsection{$\be7(n,p)\li7$}
\noindent
$N_A\langle \sigma v\rangle$ = 5.17900e9*(1.-1.44587*t912+1.12925*t9-.493526*t932
     \\    +.126269*t9*t9-.0194265*t932*t9+.00177188*t9**3
     \\    -.883411e-4*t912*t9**3+.185551e-5*t9**4) 
     \\     +4.2994e9*t9m32*ex(-3.713442/t9)
     \\     +1.36949e11*t9m32*ex(-31.332167/t9)

\subsection{$\li7(p,\alpha)\he4$}
\noindent
$N_A\langle \sigma v\rangle$ = 9.19322e8*t9m23*ex(-8.4730/t913)
     \\     *(1. - 2.26222*t913 + 11.3224*t923
     \\     - 27.3071*t9 + 41.1901*t943 - 37.4242*t953
     \\     + 18.3941*t9*t9 -3.72281*t9*t943 +2.58125e-2*t953*t9)

\end{document}